\documentclass[longauth]{aa} 

\usepackage{graphicx}
\usepackage{txfonts}
\usepackage{natbib}
\usepackage{orcidlink}
\usepackage{enumitem}
\usepackage{lscape} 
\usepackage{float}
\usepackage{multirow}
\usepackage{subcaption}
\usepackage{hyperref}
\begin{document} 

\title{The Hot Neptune Initiative (HONEI)}
\subtitle{III. An ultra-hot super Neptune orbiting the metal-rich subgiant star TOI-5646: \\ the largest host for a desert dweller}
\titlerunning{The largest host for a desert dweller}
\authorrunning{L. Mancini et al.}

\author{
L. Mancini\orcidlink{0000-0002-9428-8732}\inst{\ref{inst:tor_verg},\ref{inst:oato}}\fnmsep\thanks{Corresponding author: lmancini@roma2.infn.it}
\and
F. Manni\orcidlink{0009-0003-1841-7381}\inst{\ref{inst:tor_verg},\ref{inst:oato}}
\and
D. Kubyshkina\orcidlink{0000-0001-9137-9818}\inst{\ref{inst:sria},\ref{inst:srp}}
\and
E.~A. van Dijk\orcidlink{0009-0000-2307-2187}\inst{\ref{inst:leiden}}
\and
L. Naponiello\orcidlink{0000-0001-9390-0988}\inst{\ref{inst:oato}}
\and
M. Damasso\orcidlink{0000-0001-9984-4278}\inst{\ref{inst:oato}}
\and
V. D'Orazi\orcidlink{0000-0002-2662-3762}\inst{\ref{inst:tor_verg},\ref{inst:oar}}
\and
K. Biazzo\orcidlink{0000-0002-1892-2180}\inst{\ref{inst:oar}}
\and
A.~V. Bonomo\orcidlink{0000-0002-6177-198X}\inst{\ref{inst:oato}}
\and
L. Fossati\orcidlink{0000-0003-4426-9530}\inst{\ref{inst:sria}}
\and
Y. Miguel\orcidlink{0000-0002-0747-8862}\inst{\ref{inst:leiden},\ref{inst:sron}}
\and
J. de Leon\orcidlink{0000-0002-6424-3410}\inst{\ref{inst:komaba}}
\and
M.~E. Everett\orcidlink{0000-0002-0885-7215}\inst{\ref{inst:noirlab}}
\and
W. Fong\orcidlink{0000-0003-0241-2757}\inst{\ref{inst:kavli}}
\and
A. Fukui\orcidlink{0000-0002-4909-5763}\inst{\ref{inst:komaba},\ref{inst:canarias}}
\and
D. Gandolfi\orcidlink{0000-0001-8627-9628}\inst{\ref{inst:unito}}
\and
S.~B. Howell\orcidlink{0000-0002-2532-2853}\inst{\ref{inst:nasa}}
\and
D.~W. Latham\orcidlink{0000-0001-9911-7388}\inst{\ref{inst:harvard}}
\and
M. L\'opez-Morales\orcidlink{0000-0003-3204-8183}\inst{\ref{inst:stsi}}
\and
P. MacQueen\orcidlink{0009-0009-6344-9589}\inst{\ref{inst:mcdonald}}
\and
F. Murgas\orcidlink{0000-0001-9087-1245}\inst{\ref{inst:canarias},\ref{inst:laguna}}
\and
N. Narita\orcidlink{0000-0001-8511-2981}\inst{\ref{inst:komaba},\ref{inst:osawa},\ref{inst:canarias}}
\and
E. Palle\orcidlink{0000-0003-0987-1593}\inst{\ref{inst:canarias},\ref{inst:laguna}}
\and
M. Pinamonti\orcidlink{0000-0003-0987-1593}\inst{\ref{inst:oato}}
\and
T. Pugh\orcidlink{0009-0009-6344-9589}\inst{\ref{inst:mcdonald}} 
\and
A. Sozzetti\orcidlink{0000-0002-7504-365X}\inst{\ref{inst:oato}}
\and
S. Vissapragada\orcidlink{0000-0003-2527-1475}\inst{\ref{inst:pasadena}}
\and
S.~W. Yee\orcidlink{0000-0001-7961-3907}\inst{\ref{inst:angeles}}
}

\institute{
Department of Physics, University of Rome ``Tor Vergata'', Via della Ricerca Scientifica 1, 00133 -- Rome, Italy \label{inst:tor_verg} 
\and
INAF -- Turin Astrophysical Observatory, via Osservatorio 20, 10025 -- Pino Torinese, Italy \label{inst:oato}
\and
Space Research Institute, Austrian Academy of Sciences, Schmiedlstrasse 6, 8042 -- Graz, Austria \label{inst:sria}
\and
Space Research and Planetology, Physics Institute, University of Bern, Geselschaftsstrasse 6, 3012 -- Bern, Switzerland \label{inst:srp}
\and
Leiden Observatory, Leiden University, Einsteinweg 55, 2333 CC Leiden, the Netherlands \label{inst:leiden}
\and
INAF -- Astronomical Observatory of Rome, Via Frascati 33, 00178 -- Monte Porzio Catone (RM), Italy \label{inst:oar}
\and
SRON Netherlands Institute for Space Research, Niels Bohrweg 4, 2333 CA Leiden, the Netherlands  \label{inst:sron}
\and
Komaba Institute for Science, The University of Tokyo, 3-8-1 Komaba, Meguro, Tokyo 153-8902, Japan \label{inst:komaba}
\and
NSF NOIRLab, 950 N. Cherry Ave., Tucson, AZ 85719, USA \label{inst:noirlab}
\and
Kavli Institute for Astrophysics and Space Research, Massachusetts Institute of Technology, Cambridge, MA 02139, USA \label{inst:kavli}
\and
Instituto de Astrof\'{i}sica de Canarias (IAC), 38205 La Laguna, Tenerife, Spain \label{inst:canarias}
\and
Department of Physics, University of Turin, Via Pietro Giuria 1, 10125 -- Torino, Italy \label{inst:unito}
\and
NASA Ames Research Center, Moffett Field, CA 94035, USA \label{inst:nasa}
\and
Center for Astrophysics | Harvard \& Smithsonian, 60 Garden Street, Cambridge, Massachusetts 02138, USA \label{inst:harvard}
\and
Space Telescope Science Institute, 3700 San Martin Dr. Baltimore, MD 21218, USA \label{inst:stsi}
\and
McDonald Observatory and Center for Planetary Systems Habitability, The University of Texas, Austin, Texas, USA \label{inst:mcdonald}
\and
Departamento de Astrof\'isica, Universidad de La Laguna (ULL), E-38206 La Laguna, Tenerife, Spain \label{inst:laguna}
\and
Astrobiology Center, 2-21-1 Osawa, Mitaka, Tokyo 181-8588, Japan \label{inst:osawa}
\and
Carnegie Science Observatories, 813 Santa Barbara Street, Pasadena, CA 91101, USA \label{inst:pasadena}
\and
Department of Physics \& Astronomy, University of California Los Angeles, Los Angeles, CA 90095, USA \label{inst:angeles}
}
   \date{Received 5 June 2026 / Accepted 14 August 2026}

\abstract
{
Neptune-sized exoplanets appear to be located in three distinct regions when plotted in the period-radius space: the desert, which is an ultra-close-in region nearest to the host star and is typically defined by orbital periods shorter than 3.2 days and is characterised by a scarcity of planets; the savanna, in which planets reside at a safer distance from their hosts and are found in moderate abundance; and an intermediate region called the ridge, which features a surprising statistical overdensity. The formation and evolution pathways by which a planet ends up in one of these three regions are still poorly understood. This makes any new planet that is discovered in these regions, especially in the desert, a key new laboratory.
We report the confirmation of the TESS transiting-planet candidate orbiting the metal-rich F8\,IV-V subgiant star TOI-5646 ($V$\,=\,11.38\,mag; $T_{\rm eff}=6136 \pm 51$\,K). The planetary nature of TOI-5646\,b was confirmed by means of HARPS-N follow-up radial velocity measurements. We determined that the planet has an orbital period of $2.427702 \pm 0.000025$ days and an orbital eccentricity compatible with zero. With a mass of $45.9 \pm 4.4\,M_{\oplus}$, a radius of $6.6 \pm 0.3\,R_{\oplus}$, and an equilibrium temperature of $1935 \pm 29$ K, it is one of the hottest super-Neptune exoplanets ever discovered in the Neptune desert. Its density, $0.87 \pm 0.15$\,g\,cm$^{-3}$, is lower than that of most other desert dwellers, but still appears to follow the trend that the desert is populated by dense planets than those in the ridge and savanna. The metallicity of the parent star ([Fe/H]\,=\,$0.32 \pm 0.06$) follows the trend whereby the host stars of Neptune-sized exoplanets in the desert and the ridge are generally richer in metal than those in the savanna.
The $\log{g}$ of the parent star, $4.09 \pm 0.08$, suggests that the host is an evolved star, making this a peculiar system: TOI-5646 is the largest star ($\sim$\,$1.8\,R_{\odot}$) ever discovered to host a Neptune-sized planet in the desert. Internal structure modelling revealed that TOI-5646\,b is highly enriched in metals, with an inferred bulk metallicity between 0.7 and 0.95. Atmospheric evolution simulations indicate that the planet has undergone significant mass loss ($6-16\%$) through intense XUV irradiation from its host star. 
These results indicate a peculiar formation history (potentially involving early protoplanetary disk dispersal), a planetary collision, or a late tidal destruction event, to explain its current state as a low-density survivor in the Neptune desert.}


   \maketitle
%

\section{Introduction}
\label{sec:introduction}
Intermediate-mass giant exoplanets, often referred to as super-Neptunes or sub-Saturns, bridge the gap between ice and gas giants. These planets occupy a vast region of the parameter space that is unrepresented in our Solar System. In particular, those on ultra-short orbits ($P_{\rm orb} < 3$ days) inhabit the so-called Neptune desert. This region of low occurrence remains a theoretical puzzle, because these planets are rarer than smaller super-Earths and larger gas giants, even though they are relatively easy to detect. Although early data from the {\it Kepler} space telescope suggested that the desert was empty \citep{latham2011}, the Transiting Exoplanet Survey Satellite (TESS) \citep{ricker2015} has since identified several dwellers, revealing a complex population potentially shaped by photoevaporation and high-eccentricity migration \citep{mazeh2016,owen2019}. These survivors show a surprising diversity in density: some ultra-hot Neptunes retain thick envelopes but have low core masses \citep{jenkins2020,nabbie2024,grunblatt2024}, while others have exceptionally dense cores \citep{armstrong2020,osborn2023} that may have formed through catastrophic planetary collisions \citep{naponiello2023}. 
Overall, the upper edge of the Neptune desert is populated by massive ($M_{\rm p}>10\,M_{\oplus}$), dense ($\rho_{\rm p}\gtrsim  1$\,g\,cm$^{-3}$) planets that likely survived intense stellar X-ray and extreme ultraviolet radiation; conversely, this high-energy radiation effectively eroded the atmospheres of planets located along the lower boundary of the desert \citep{owen2018}. 
Because these massive Neptunes orbit high-metallicity hosts that are indistinguishable from those of hot Jupiters, they are hypothesised to be the remnant cores of stripped hot Jupiters \citep{vissapragada2025,hallatt2026}.
Beyond the Neptune desert, the distribution of Neptune-sized exoplanets shifts into the ridge and savanna zones (\citealt{castro2024a}; see Fig.~\ref{fig:nep_des}), each defined by physical and orbital trends that are still remain to be fully understood but appear to be distinct from those of desert dwellers
\citep[e.g.][]{vissapragada2025,manni2025,mancini2026}.

As part of the long-term Global Architecture of Planetary Systems (GAPS; see, e.g., \citealt{damasso2015,esposito2017}) project, we ran an observational programme to confirm and characterise intermediate-mass planets with sizes spanning the regime of super-Earth to sub-Saturn (2\,$R_{\oplus}$\,<\,$R_{\rm p}$<\,9\,$R_{\oplus}$) \citep{naponiello2022,naponiello2023,naponiello2025a,naponiello2026a}. This effort was later followed by the more focused Hot Neptune Initiative (HONEI) programme \citep{naponiello2025b,manni2025}, which aims to confirm Neptune-sized candidates suitable for atmospheric observations. The two initiatives primarily rely on data from the TESS survey (Sect.~\ref{sec:TESSdata}) and the High Accuracy Radial velocity Planet Searcher for the Northern hemisphere (HARPS-N) spectrograph (Sect.~\ref{sec:harpsn}). The selection of the HONEI targets was described by \citet{naponiello2025b} and is essentially based on the full list of TESS objects of interest (TOIs) with estimated sizes of 3\,$R_{\oplus}$\,<\,$R_{\rm p}$<\,7\,$R_{\oplus}$ within $1\,\sigma$ uncertainties on $R_{\rm p}$. 
In this latest paper of our series, we present new TESS photometry and HARPS-N radial velocity (RV) measurements, the analysis of which forms the basis of this work, to confirm the planetary nature of TOI-5646\,b and to fully characterise the physical and orbital parameters of the TOI-5646 planetary system, which is located $\sim$\,400\,pc away from us.

The paper is organised as follows. 
In Sect.~\ref{sec:data} we describe the TESS photometry and present new time series of HARPS-N data. The analysis of these data is the basis this work. Ground-based photometry and imaging data with a high-angular-resolution of TOI-5646 are also presented.
In Sect.~\ref{sec:host_star} we characterise the parent star and we characterise its planet in Sect.~\ref{sec:analysis_characterisation}. 
The properties of TOI-5646\,b are discussed in Sect.~\ref{sec:discussion} in the context of the Neptune desert and the more populated ridge and savanna regions. In this section, we also investigate the internal composition of TOI-5646\,b and its atmospheric evolution and formation scenario.
Finally, a summary of our results is given in Sect.~\ref{sec:conclusions}.
%

\section{Observations and data reduction}
\label{sec:data}

\subsection{TESS photometry}
\label{sec:TESSdata}

The star TIC\,366804698 (aka 2MASS\,J13134732+0542077) was first observed by TESS in March 2020 in Sector 23, with an exposure time ($T_{\rm exp}$) of 1800\,s over a 26-day baseline. Observations continued in Sectors 46 and 50 with a shorter cadence of $T_{\rm exp}$\,=\,600\,s for an additional 54 days. After it was identified as a TESS Object of Interest (TOI), the target was catalogued as TOI-5646 on June 2, 2022. It was subsequently re-observed in April 2025 in Sector 91 for 18 days in short-cadence mode ($T_{\rm exp}$\,=\,120\,s), and a total of more than 23\,500 images were accumulated. The details of these TESS observations are summarised in Table~\ref{tab:TESS_obs}.

%
\begin{table}
\centering %
\caption{Details of the TESS observations of TOI-5646.}
\label{tab:TESS_obs}
\begin{tabular}{crrcc}
\hline %
\hline  \\[-8pt]
Sector & $T_{\rm exp}$\,(s)& $N_{\rm obs}$ & Start & End \\ [2 pt]
\hline  \\[-6pt] %
23 & 1800~~~ & 1020  & 2020-03-20 & 2020-04-15 \\
46 &  600~~~ & 3558  & 2021-12-03 & 2021-12-30 \\
50 &  600~~~ & 3003  & 2022-03-26 & 2022-04-22 \\
91 &  120~~~ & 16011 & 2025-04-09 & 2025-05-07 \\
\hline %
\end{tabular}
\tablefoot{$T_{\rm exp}$ and $N_{\rm obs}$ are the exposure time adopted and the number of observations obtained for each sector, respectively.}
\end{table}

For each sector, we downloaded the TOI-5646 TESS light curves from the Mikulski Archive for Space Telescopes (MAST) using the Python package \texttt{lightkurve} \citep{lightkurve}. The fluxes were extracted via the TESS Science Processing Operations Center (SPOC) pipeline  \citep{jenkins2016}, using the Presearch Data Conditioning Simple Aperture Photometry (PDC-SAP; \citealt{Stumpe2012,Stumpe2014,Smith2012}), which also takes the potential contamination from nearby stars into account. The default aperture of the SPOC pipeline was adopted. These PDC-SAP light curves are plotted in Fig.~\ref{fig:TESS_lc}. In Sectors 23, 46, 50, and 91, we clearly identified 8, 10, 7, and 5 transit events of the planet candidate TOI\,5646.01, respectively, with a well-defined periodicity of $\sim$\,2.43\,days.

\subsection{Search for contaminant sources in TESS photometry}
Because TESS has very large pixels ($\sim$\,21\,arcsec), the light from nearby stars or background objects frequently blends with that of the target star, potentially leading to a transit depth that is shallower than the actual value. It is therefore a critical diagnostic step to check for contaminating sources within the TESS aperture.
We used the Python package \texttt{TESS-cont} to search for nearby Gaia DR2 \citep{gaia2018} or DR3 \citep{gaia2023} sources and to quantify their flux contribution to the TESS photometry \citep{castro2024}. The corresponding heatmap (see Fig.~\ref{fig:TESS-cont}) shows that TOI-5646 contributes no less than 98\% of the total flux in every pixel of the SPOC aperture in all four Sectors (23, 46, 50, and 91) containing target data.

\subsection{MuSCAT2 time-series photometry}
Using public ephemerides, we observed a partial and a full transit of TOI\,5646.01 on February 18 and March 7, 2026, respectively, both under clear sky conditions. Observations were simultaneously obtained in the $g$, $r$, $i$ and $z_{\rm s}$ bands using the MuSCAT2 \citep{narita2019} instrument mounted on the 1.52\,m TCS telescope at the Teide Observatory on the Canary Islands (Spain). MuSCAT2 enables simultaneous imaging in four photometric bands, each with a $1\mathrm{k}\times1\mathrm{k}$ CCD camera featuring a pixel scale of $0.^{\prime \prime}44$\,pixel$^{-1}$ and a field of view (FOV) of $7.^{\prime}4 \times 7.^{\prime}4$. Exposure times ranged from 3 to 5\,s depending on the band. A slight telescope defocus was applied to improve the photometric precision (e.g. \citealt{southworth2012,southworth2015,mancini2013}), while avoiding flux contamination by a nearby star located $\sim$\,$13^{\prime \prime}$ from the target.
Photometry was performed using the MuSCAT2 pipeline\footnote{\url{https://github.com/hpparvi/MuSCAT2\_transit\_pipeline}} 
\citep{parviainen2020,parviainen2022}. Briefly, the pipeline computes aperture photometry using multiple comparison stars and aperture sizes; it then generates the final relative light curves through a global optimisation of the posterior distribution. This optimisation incorporates a comprehensive model that includes aperture and comparison-star selection, the transit profile, and a linear baseline component, treating airmass, centroid shifts (in the $x$ and $y$ directions), and sky background level as covariates. Preliminary independent fits for each epoch indicated that the transit depths exhibit no significant chromaticity, although an apparent discrepancy between the $g$ and $z$ band depths likely stems from the lack of a pre-ingress baseline. The final MuSCAT2 light curves are shown in Fig.~\ref{fig:muscat}.

\subsection{MuSCAT4 time-series photometry}
Another transit of TOI\,5646.01 was observed with the MuSCAT4 camera \citep{narita2020} mounted on the LCOGT\,2\,m telescope at the Siding Spring Observatory on April 8, 2026, through a filter set similar to the one used for MuSCAT2 \citep{fukui2024}. However, the quality of the photometry extracted from these four-colour data proved too low to be of practical use. Therefore, we did not consider these data in our subsequent characterisation of the TOI-5646 system.

\subsection{Imaging with a high angular resolution }
When the host star of an exoplanet harbours a spatially close companion, this companion (regardless of whether physically bound or aligned along the line of sight) can create a false-positive transit signal when it is, for example, an eclipsing binary (EB) or another type of variable star. Furthermore, third-light flux from such a close companion star will lead to an underestimated planetary radius and incorrect physical parameters when it is not accounted for in the transit model \citep{ciardi2015,furlan2017,furlan2020,southworth2020,mancini2022}. To search for close-in bound companions that are unresolved in the TESS images, we therefore obtained high-resolution speckle-imaging observations of TOI-5646 to supplement the Gaia data. 
\subsubsection{NESSI at the  WIYN 3.5\,m telescope}
We observed TOI-5646 on February 5, 2023, using the NN-EXPLORE Exoplanet Stellar Speckle Imager (NESSI; \citealt{scott2018}) at the WIYN 3.5\,m telescope at the Kitt Peak National Observatory. The data consist of 9000 40\,ms frames acquired in two filters with central wavelengths at $\lambda_{\rm c}$\,=\,562 and 832~nm. The FOV of NESSI was confined to a $256\times256$ pixel sub-array readout ($4.6\arcsec\times4.6\arcsec$). However, our speckle measurements were further restricted to an outer radius of $1.2\arcsec$ from the target star. A similar set of 1000 speckle frames was acquired for a nearby single star to calibrate the PSF of the TOI-5646 data.
These speckle data were reduced using the pipeline described by \citet{howell2011}. The pipeline products include a reconstructed image of the field surrounding the target in each filter. We obtained a contrast curve from each reconstructed image by measuring fluctuations in the noise-like background level as a function of angular separation from TOI-5646 (see the left panel of Fig.~\ref{fig:hri}). These contrast curves establish upper limits on the relative brightness of any undetected point source in close proximity to the target star. No companion sources for TOI-5646 were detected in the NESSI data.
\subsubsection{‘Alopeke at the Gemini North 8\,m telescope}
TOI-5646 was observed on UT February 18, 2025, using the ‘Alopeke speckle instrument on the Gemini North 8\,m telescope \citep{scott2021}. ‘Alopeke provides simultaneous speckle imaging in two bands (562 and 832 nm) with output data products that include a reconstructed image with robust contrast limits for companion detections. Five sets of 1000\,$\times$\,0.06\,s exposures were collected and subjected to Fourier analysis in our standard reduction pipeline (see \citealt{howell2011}). The right panel of Fig.~\ref{fig:hri} shows our final 5\,$\sigma$ magnitude contrast curves and the 832\,nm reconstructed speckle image. TOI-5646 is a single star without a companion brighter than five to eight magnitudes below that of the target star from the diffraction limit (20\,mas) out to 1.2$^{\prime \prime}$. At the distance of TOI-5646 ($d$\,=\,429\,pc), these angular limits correspond to spatial limits of 8.6 to 515\,au.

\subsection{TRES data}
\label{sec:tres}
Two  reconnaissance spectra near opposite quadratures were obtained in June 2022 with the Tillinghast Reflector Echelle Spectrograph (TRES; \citealt{szentgyorgyi2007}), which is a fiber-fed echelle spectrograph ($R$\,=\,44\,000) mounted on the 1.5\,m Tillinghast telescope at the Smithsonian Astrophysical Observatory's Fred L. Whipple Observatory on Mount Hopkins in Arizona (USA). Three more observations were collected in February 2025 at the request of our group. The multi-order relative velocities from the five TRES observations are consistent with a circular orbit constrained by the TESS ephemeris, yielding a semi-amplitude of 19\,m\,s$^{-1}$ and rms residuals of about 15\,m\,s$^{-1}$. The SPC (Stellar Parameter Classification) automated analysis pipeline, used to derive stellar properties from TRES echelle spectra, yielded the following stellar parameters: $T_{\rm eff}$\,=\,$6087 \pm 50$\,K, $\log{g}$\,=\,$4.22 \pm 0.10$, ${\rm [Fe/H]}$\,=\,$+0.36 \pm 0.08$, and $v_{\rm rot}$\,=\,$4.8 \pm 0.5$\,km\,s$^{-1}$ (not corrected for macroturbulence), which agree well with the final values we adopted (Table~\ref{tab:star_TOI-5646}).

\subsection{Tull Coud\'e spectrograph data}
\label{sec:tull}
We acquired three additional high-resolution reconnaissance spectra of TOI-5646 using the Tull Coud\'e spectrograph, which is mounted on the 2.7\,m Harlan J. Smith Telescope at the McDonald Observatory in Texas, USA \citep{tull1995}. We used the TS23 mode, which provides a resolving power of $R$\,$\approx$\,$60\,000$ over the $375-1000$\,nm wavelength range. We placed a temperature-stabilised I$_2$ gas cell in the light beam to trace instrumental drift and enable high-precision RV measurements. We reduced the spectra using standard IRAF routines and extracted the RVs with the code \textit{Austral} \citep{endl2000}. The Doppler measurements do not exhibit significant variation when phased with the TESS transit ephemeris, producing a semi-amplitude of $K$\,=\,$11_{-8}^{+10}$\,m\,s$^{-1}$ with a mean uncertainty of 15\,m\,s$^{-1}$. This rules out an eclipsing-binary scenario for the target star.

\subsection{HARPS-N data}
\label{sec:harpsn}
\begin{figure*}
\centering
\includegraphics[width=0.9\textwidth]{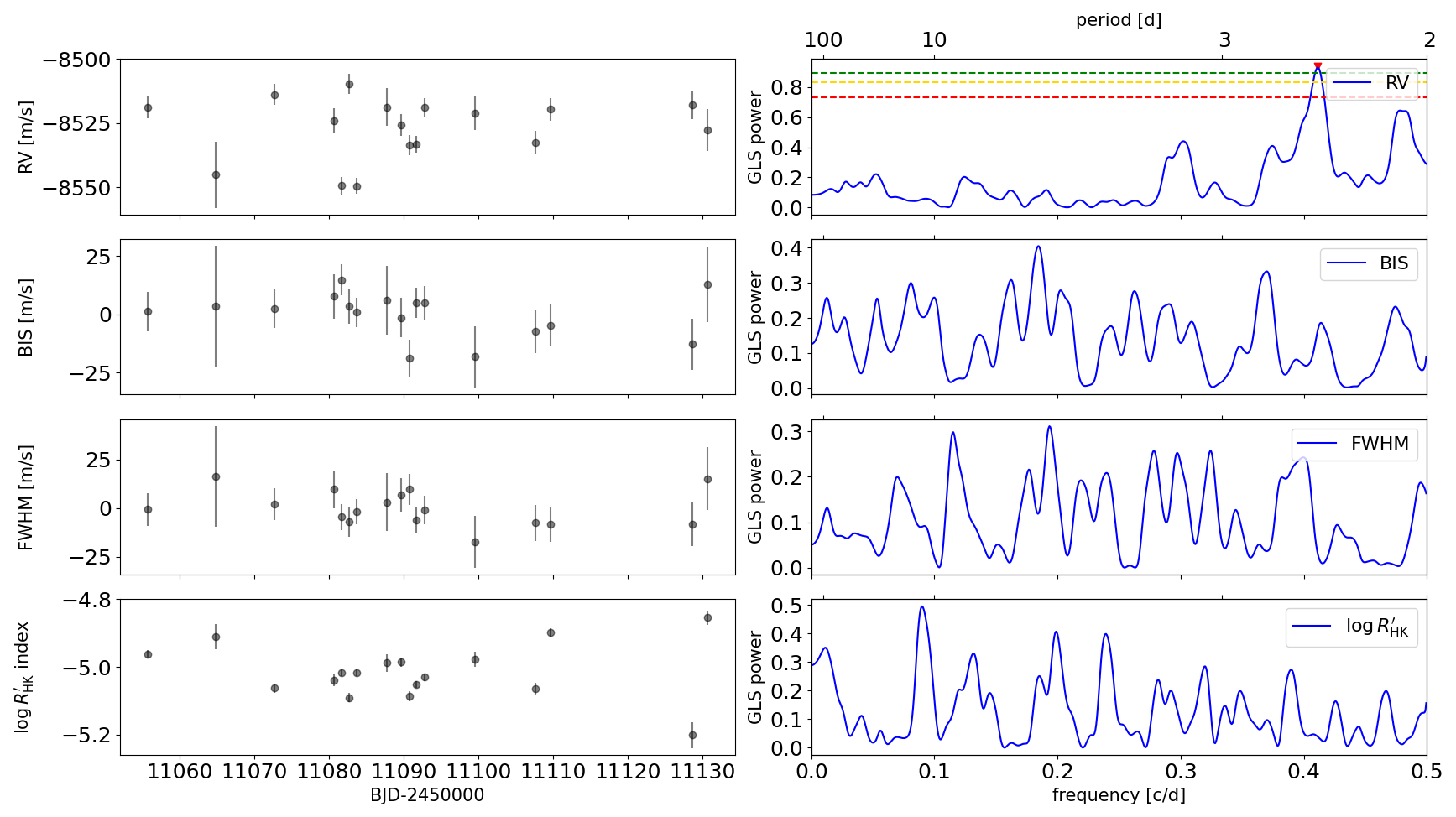}
\caption{Time series (left panels) and GLS periodograms (right panels) of RVs and activity diagnostics. The red triangle marks the orbital period of planet $b$. The levels of FAP, determined through a bootstrap analysis, are indicated by dashed horizontal lines (green: FAP\,=\,0.1\%; yellow: FAP\,=\,1\%; red: FAP\,=\,10\%). }
\label{fig:rvactvity}
\end{figure*}

Mounted on the 3.58\,m Italian Telescopio Nazionale Galileo (TNG) at the Observatorio del Roque de los Muchachos in La Palma (Spain), HARPS-N is a high-resolution ($R$\,=\,115\,000) visible-light (383\,nm\,$\leq$\,$\lambda$\,$\leq$\,690\,nm) fiber-fed echelle spectrograph. It remains a premier instrument for RV studies, capable of reaching $\sim$\,1\,m\,s$^{-1}$ precision (under optimal observing conditions) through its exceptional long-term stability and simultaneous wavelength calibration \citep{cosentino2012}.
We used HARPS-N to obtain precise RV measurements of TOI-5646 from January 15 to March 31, 2026, as part of the programme ${\rm A52TAC\_38}$ (PI: F. Manni). The exposure time was selected between 15 and 30 minutes, depending on the weather conditions. The fiber AB spectroscopy mode (object plus sky) was adopted. We collected a total of 17 RV measurements with an average uncertainty of 5.2\,m\,s$^{-1}$ and an average signal-to-noise ratio (S/N) of 28.4 at 550\,nm (see the upper panel of Fig.~\ref{fig:rvactvity} and Table~\ref{tab:RV_TOI-5646}).

\section{Host-star characterisation}
\label{sec:host_star}

\subsection{Atmospheric parameters and iron abundance}
\label{subsec:host_star_atmos}
We employed the code ARES \citep{sousa2015} to measure the equivalent widths (EWs) of the \ion{Fe}{i} and \ion{Fe}{ii} lines. We used the same line list in \citet{dalponte2025}. These EW values were then fed into the $q^2$ code, a Python wrapper for MOOG \citep{sneden1973}, which was developed by \citet{ramirez2014}. We used ATLAS models \citep{castelli2003} with {\tt ODFNEW} opacities. The derived stellar parameters and iron abundance are presented in Table \ref{tab:star_TOI-5646}. For further details regarding abundance determination and error analysis, we refer to our previous publications (e.g. \citealt{naponiello2025b,manni2025}).
Based on the derived stellar atmospheric parameters, we determined the projected rotational velocity ($v \sin{i_{\star}}$) via spectral synthesis using the {\it synth} driver of the MOOG code. Following the method of \citet{Biazzoetal2022}, we adopted a fixed macroturbulence velocity ($v_{\rm macro}$) of 5.0\,km\,s$^{-1}$ (\citealt{doyle2014}). This yielded a value of $v \sin{i_{\star}}$\,=\,2.0\,$\pm$\,0.8\,km\,s$^{-1}$, which agrees excellently with the value of 2.0\,$\pm$\,0.7\,km\,s$^{-1}$ obtained by applying the cross-correlation function (CCF) and the full width at half maximum (FWHM) calibrations from \citet{Raineretal2023}. 
Furthermore, we identified the lithium resonance line at $\sim$\,6707.8\,\AA\, in the same co-added spectrum used for the atmospheric parameters. We measured a lithium EW of 44.5\,$\pm$\,2.5\,m\AA, which translates into an abundance of $\log n{\rm (Li)}_{\rm NLTE}$\,=\,2.54\,$\pm$\,0.04\,dex after applying the non-local thermodynamic equilibrium (NLTE) corrections provided by \citet{Lindetal2009}. This lithium abundance is consistent with the levels observed in open clusters with ages of around $2-4$\,Gyr (\citealt{SestitoRandich2005}). By employing the code {\sc eagles} and following the approach of \citet{jeffriesetal2023}, we derived an age estimate of $1.8^{+3.2}_{-1.6}$ Gyr based on the lithium EW and the effective temperature. This value agrees within uncertainties with the value estimated by the analysis described in the next section.

\subsection{Stellar parameters}
To derive the stellar mass, radius, and age, we simultaneously modelled the stellar spectral energy distribution (SED) and the MIST evolutionary tracks (e.g. \citealt{Paxton2015}) in a Bayesian differential evolution Markov chain Monte Carlo framework with {\tt EXOFASTv2} (\citealt{2017ascl.soft10003E, Eastman2019}; see also \citealt{naponiello2025a} for more details). We sampled the SED with the optical APASS Johnson $B$, $V$ and Sloan $g'$, $i'$ magnitudes, the near-infrared 2MASS $J$, $H$, and $K_{\rm s}$ magnitudes, and the infrared WISE $W1$, $W2$, and $W3$ magnitudes (see Table~\ref{tab:star_TOI-5646} and Fig.~\ref{fig:SED}). We imposed Gaussian priors on the $T_{\rm eff}$ and [Fe/H] derived from the analysis of the HARPS-N spectra (Sect.~\ref{subsec:host_star_atmos}) as well as on the Gaia DR3 parallax. We then computed the values and $1\sigma$ uncertainties of the stellar parameters as the medians and 15.86\%\,$-$\,84.14\% percentiles of their posterior distributions, respectively. We found $M_\star=1.418^{+0.068}_{-0.079} \, M_\odot$, $R_\star=1.837^{+0.075}_{-0.066} \, R_\odot$, and  $t_{\rm age} = 2.82^{+1.00}_{-0.69}$~Gyr (see Table~\ref{tab:star_TOI-5646}), indicating that the host is an F8\,IV-V star leaving the main sequence and becoming a sub-giant. The surface gravity we derived is fully consistent with the spectroscopic value (cf. Table~\ref{tab:star_TOI-5646}). 

\subsection{Activity indices}
We examined a set of spectroscopic activity diagnostics (see Table~\ref{tab:RV_TOI-5646}) to characterise the activity levels of TOI-5646: the CCF-based indicator bisector inverse slope (BIS), the FWHM, and the chromospheric index $\log R^{\rm \prime}_{\rm HK}$. The BIS and FWHM were calculated by the DRS pipeline, and the $\log R^{\rm \prime}_{\rm HK}$ index was determined using the publicly available code \texttt{ACTIN2}\footnote{\url{https://actin2.readthedocs.io/en/latest/index.html}} \citep{gomesdasilva2018JOSS....3..667G,gomesdasilva2021A&A...646A..77G}. The time series and the generalized Lomb-Scargle (GLS; \citealt{zechmeister2009}) periodograms are shown in Fig.~\ref{fig:rvactvity}. None of the activity diagnostics reveals significant periodicity, and the mean value of the $\log R^{\rm \prime}_{\rm HK}$ index ($-5.01$\,dex) indicates a low level of chromospheric activity.
%

\begin{table}
\centering %
\caption{Stellar parameters of TOI-5646.} %
\label{tab:star_TOI-5646} %
\resizebox{\hsize}{!}{
\begin{tabular}{lccc}
\hline %
\hline  \\[-8pt]
Parameter & Unit & Value & Source\\
 &  &  &\\
\hline  \\[-6pt] %
\multicolumn{1}{l}{\large{{\bf Cross-identifications}}} \\ [2pt] %
TOI \dotfill & \dotfill & 5646 & TOI\\
TIC ID \dotfill & \dotfill & 366804698 & TESS\\
TYC \dotfill & \dotfill & 0305-00246-1 & Tycho\\
2MASS \dotfill & \dotfill & {\tiny J13134732+0542077} & 2MASS \\
APASS \dotfill & \dotfill & 39236245 & APASS \\
UCAC4 \dotfill & \dotfill & 479-052066 & UCAC4 \\
Gaia \dotfill & \dotfill& {\tiny 3717138824247066112} & Gaia~DR3 \\ [6pt] %
\multicolumn{1}{l}{\large{{\bf Astrometric properties}}} \\ [2pt] %
$\alpha$\,(J2015.5) \dotfill & h:m:s & \,\,\,\,13:13:47.31 & Gaia~DR3 \\
$\delta$\,(J2015.5)  \dotfill & $^{\circ}$:$^{\prime}$:$^{\prime \prime}$ & +05:42:07.67 & Gaia~DR3 \\
$\pi$ \dotfill & mas & $2.4119 \pm 0.0317$ & Gaia~DR3 \\
$\mu_\alpha \cos{\delta}$ \dotfill & mas/yr  & $-15.9165 \pm 0.0361$~~~~ & Gaia~DR3 \\
$\mu_\delta$ \dotfill & mas/yr  & $~~-0.77596 \pm 0.0284$~~~~~~ & Gaia~DR3 \\ [6pt] %
\multicolumn{1}{l}{\large{{\bf Photometric properties}}} \\ [2pt] %
$B_{\rm T}$ \dotfill & mag & $12.32 \pm 0.28$ & TIC\,v8.2 \\  
$V_{\rm T}$ \dotfill & mag & $11.382 \pm 0.023$ & TIC\,v8.2 \\ 
$u$ \dotfill & mag & $14.6689 \pm 0.0062$ & TIC\,v8.2 \\  
$g$ \dotfill & mag & $11.7789 \pm 0.0008$ & TIC\,v8.2 \\  
$r$ \dotfill & mag & $11.3572 \pm 0.0008$ & TIC\,v8.2 \\ 
$i$ \dotfill & mag & $11.2334 \pm 0.0007$ & TIC\,v8.2 \\ 
$z$ \dotfill & mag & $13.5680 \pm 0.0154$ & TIC\,v8.2 \\ 
$TESS$ \dotfill & mag & $10.8829 \pm 0.0068$ & TIC\,v8.2 \\
$G$ \dotfill & mag & $11.2953 \pm 0.0004$ & Gaia~DR3 \\
$G_{\rm BP}$ \dotfill & mag & $11.5994 \pm 0.0008$ & Gaia~DR3 \\
$G_{\rm RP}$ \dotfill & mag & $10.8299 \pm 0.0003$ & Gaia~DR3 \\
$J$ \dotfill & mag & $10.328 \pm 0.023$ & 2MASS \\
$H$ \dotfill & mag & $10.050 \pm 0.023$ & 2MASS \\
$K$ \dotfill & mag & $ ~~9.989 \pm 0.021$ & 2MASS \\
$W1$\,(3.4\,$\mu$m) \dotfill & mag & $~~9.956 \pm 0.023$ & AllWISE \\
$W2$\,(4.6\,$\mu$m) \dotfill & mag & $10.010 \pm 0.019$ & AllWISE \\
$W3$\,(12\,$\mu$m)  \dotfill & mag & $10.030 \pm 0.060$ & AllWISE \\ 
$A_{V}$  \dotfill & mag & $0.156^{+0.110}_{-0.092}$ & This work \\ [6pt] 
\multicolumn{1}{l}{\large{{\bf Spectroscopic properties}}} \\ [2pt] %
Spectral type$^{(a)}$ \dotfill &  \dotfill & F8\,IV-V & This work \\
$T_{\rm eff}$ \dotfill & K & $6136 \pm 51$\,\,\,\,\,\, & This work \\
$\log g_{\star}$ \dotfill & cgs & $4.09 \pm 0.08$ & This work\\
$v\sin{i_{\star}}$ \dotfill & km\,s$^{-1}$ & $2.0 \pm 0.7$ & This work \\
$v_{\rm micro}$ \dotfill & km\,s$^{-1}$& $1.11 \pm 0.05$ & This work \\%
$v_{\rm macro}$$^{(b)}$ \dotfill & km\,s$^{-1}$& $5.0$ & This work \\%
${EW_{\rm Li}}$$^{(c)}$\dotfill & m\AA & $44.5 \pm 2.5$\,\,\, & This work \\ %
$\log{n({\rm Li})}$$^{(s)}$ \dotfill & dex & $2.54 \pm 0.04$ & This work \\ %
$\rm{[Fe/H]}$ \dotfill & dex & $+0.32 \pm 0.06$~\,\, & This work \\ [6pt] %
\multicolumn{1}{l}{\large{{\bf Derived parameters}}} \\ [2pt] %
$L_{\star}$ \dotfill & $L_{\sun}$ & $4.26^{+0.38}_{-0.29}$ & This work \\ [2pt] %
$M_{\star}$ \dotfill & $M_{\sun}$ & $1.418^{+0.068}_{-0.079}$ & This work \\ [2pt]  %
$R_{\star}$ \dotfill & $R_{\sun}$ & $1.837^{+0.075}_{-0.066}$ & This work \\ [2pt] %
$\log g_{\star}$ \dotfill & cgs & $4.060^{+0.034}_{-0.040}$ & This work \\ [2pt] %
$\rho_{\star}$\dotfill & g\,cm$^{-3}$ & $0.321^{+0.038}_{-0.038}$ & This work \\ [2pt] %
$\log R^{\prime}_{\rm HK}$\dotfill & dex &  $-5.0133 \pm 0.0035$ & This work \\ [2pt] %
$T_{\rm eff}$ \dotfill & K & $6122^{+64}_{-61}$ & This work  \\ [2pt] %
Age\dotfill & Gyr & $2.82^{+1.00}_{-0.69}$ & This work\\ [2pt] %
$A_V$ \dotfill & mag & $0.156^{+0.110}_{-0.092}$ & This work \\ [2pt] %
Distance \dotfill & pc  & $414.7^{+5.5}_{-5.3}$ & This work \\ [2pt] %
\hline %
\end{tabular}
}
\tablefoot{$^{(a)}$The spectral type was obtained from \citet{pecaut2013} (Table 5, Version 2022.04.16).
$^{(b)}$The macroturbulence velocity was computed from empirical relationships taken from \citet{doyle2014}.
$^{(c)}$This is the lithium equivalent width.
$^{(d)}$This is the NLTE lithium abundance, which was calculated by applying corrections for deviations from the Local-Thermodynamic-Equilibrium assumption. 
}
\end{table}

\section{Physical and orbital characterisation of the planetary system}
\label{sec:analysis_characterisation}

\subsection{Periodogram of the RV time series}
The HARPS-N RVs of TOI-5646 were analysed via a GLS periodogram to search for the same periodic signal identified in the TESS photometry. The highest peak in the periodogram (see the top right-hand panel of Fig.~\ref{fig:rvactvity}) clearly identifies the orbital period of the planet candidate TOI\,5646.01 as $\sim$\,2.43\,days ($f$\,$\sim$\,0.41). 

Due to the sparse RV sampling, identifying further signals remains challenging; indeed, the GLS periodogram of the one-planet model residuals shows no significant periodicities, and the TESS light curve reveals no additional transits.

\subsection{Detection sensitivity}
To quantify our sensitivity to other planetary companions, we conducted an injection-recovery analysis. Detection probabilities were computed over a logarithmic grid in planetary mass ($M_{\rm p}\sin{i}$), ranging from $10$ to $1000\,M_{\oplus}$, and semi-major axis, from $0.01$ to $10$ au. For each grid cell, 200 synthetic signals were generated with orbital parameters drawn from appropriate distributions: periods derived from Kepler’s third law, uniformly distributed times of conjunction and arguments of periastron, inclinations uniform in $\cos{i}$, and eccentricities following a Beta distribution. Planet detectability was assessed via Bayesian Information Criterion (BIC) model comparison. Synthetic RV time series were generated at the observed epochs, including Gaussian noise scaled to the instrumental jitter. Each dataset was fitted with constant, polynomial (linear or quadratic), and full Keplerian models. A signal was considered detected when the Keplerian model was strongly favoured ($\Delta\mathrm{BIC} > 10$) over simpler models; long-period companions producing significant trends were also identified when a polynomial model was preferred over a constant one.

\begin{figure}
\centering
\includegraphics[width=9.0cm]{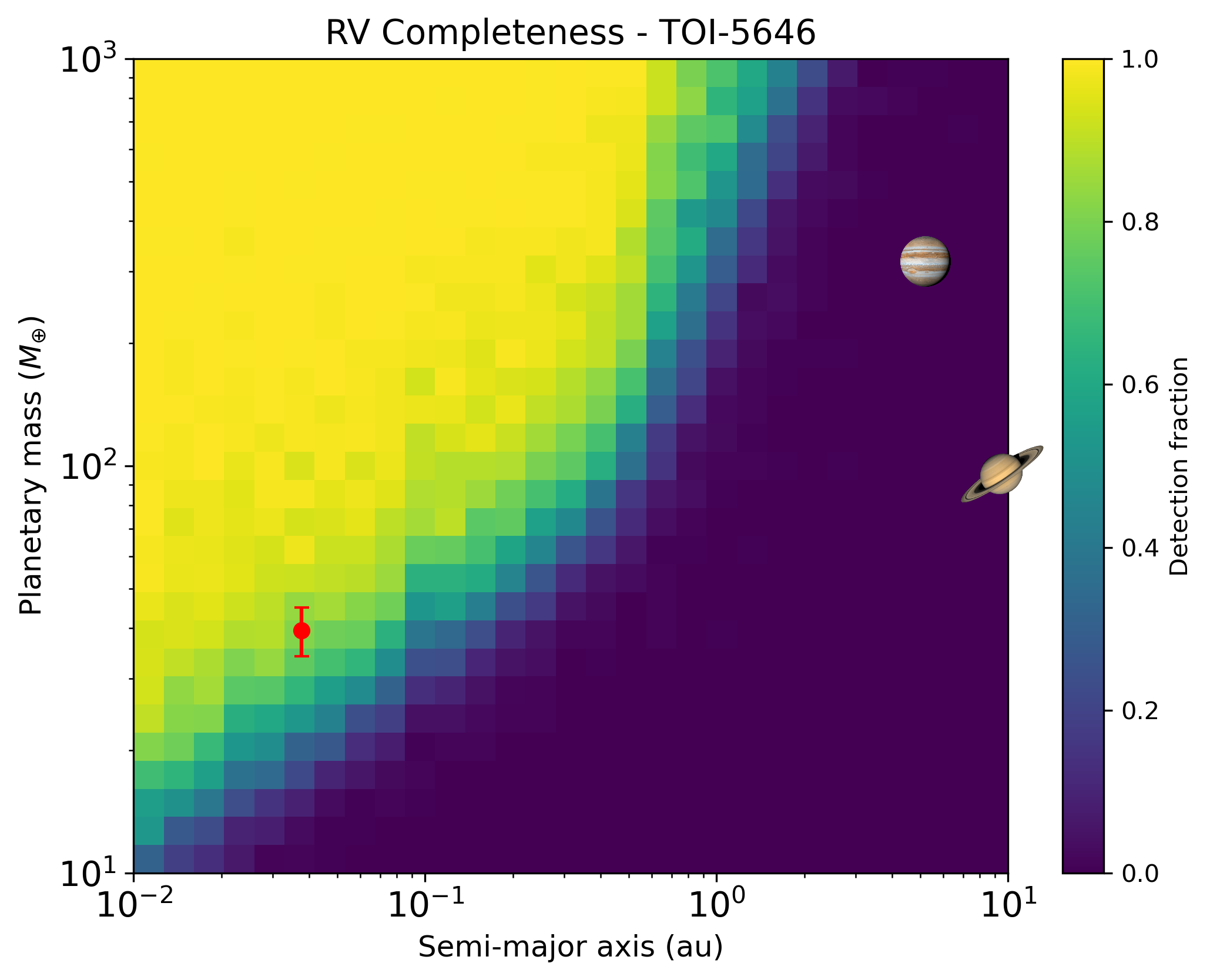}
\caption{Detection completeness map for TOI-5646 in the mass-semi-major axis plane. The colour scale indicates the recovery fraction of simulated companions. The red circle denotes the location of TOI-5646\,b, while Jupiter and Saturn are shown for reference.}
\label{fig:sensitivity}
\end{figure}

The resulting completeness map reflects the recovery fraction across the parameter space, defined as the ratio of detected to injected signals in each grid cell. As can be appreciated in Fig.~\ref{fig:sensitivity}, the current HARPS-N dataset is sensitive to Jupiter-mass companions within $\sim$\,0.5\,au, or to Saturn-mass companions within $\sim$\,0.2\,au.

\subsection{Global modelling of the data}
\label{sec:modelling}
\begin{figure}
\centering
\includegraphics[width=9.0cm]{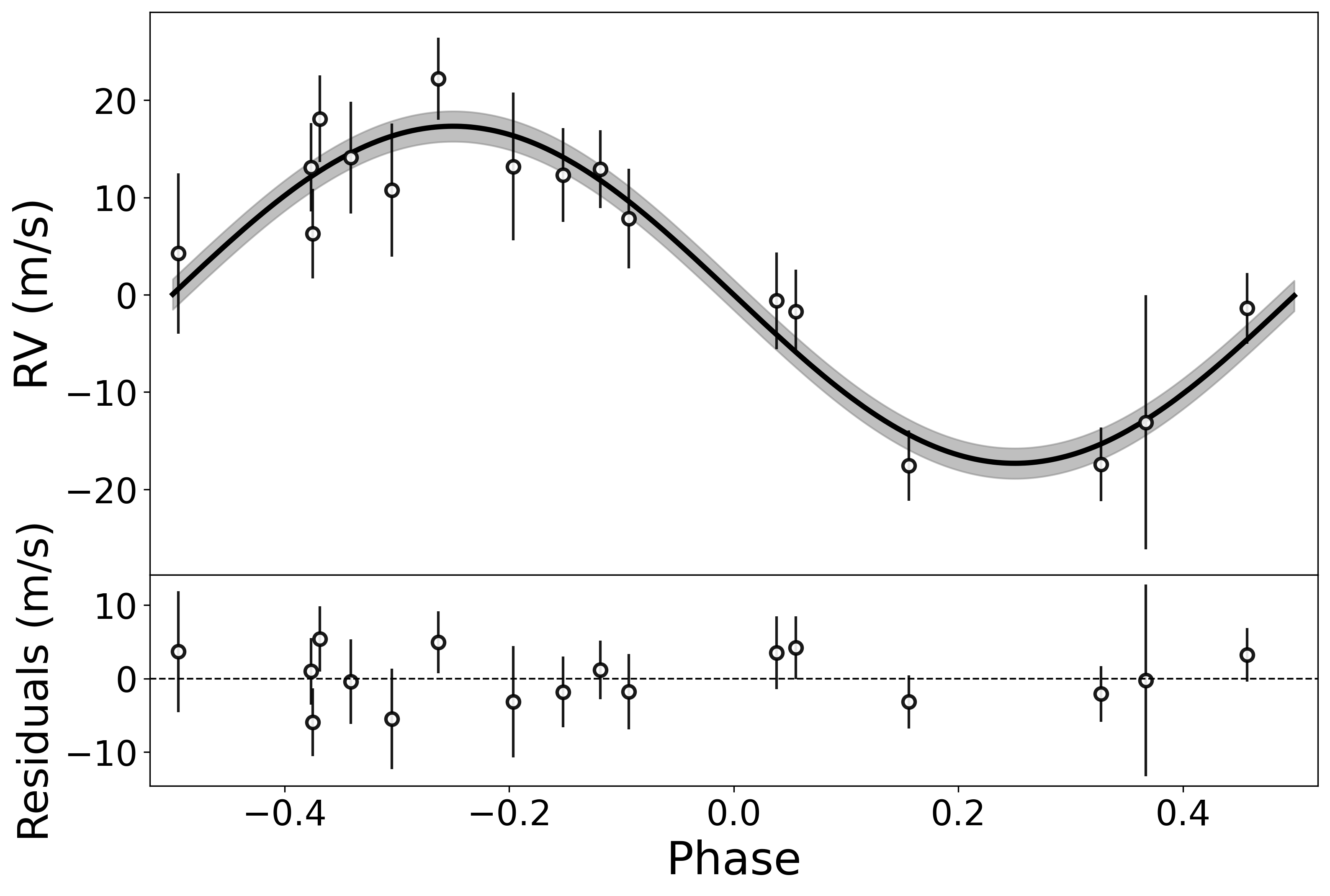}
\caption{HARPS-N RV measurements of TOI-5646 phase-folded to the period of planet TOI-5646\,b, together with our best-fit model. The error bars include both the data uncertainty and the jitter derived from the analysis. RV residuals from the best fit are shown in the bottom panel.} 
\label{fig:HARPS-N_rv_fit_phased}
\end{figure}

\begin{figure}
\centering
\includegraphics[width=9.0cm]{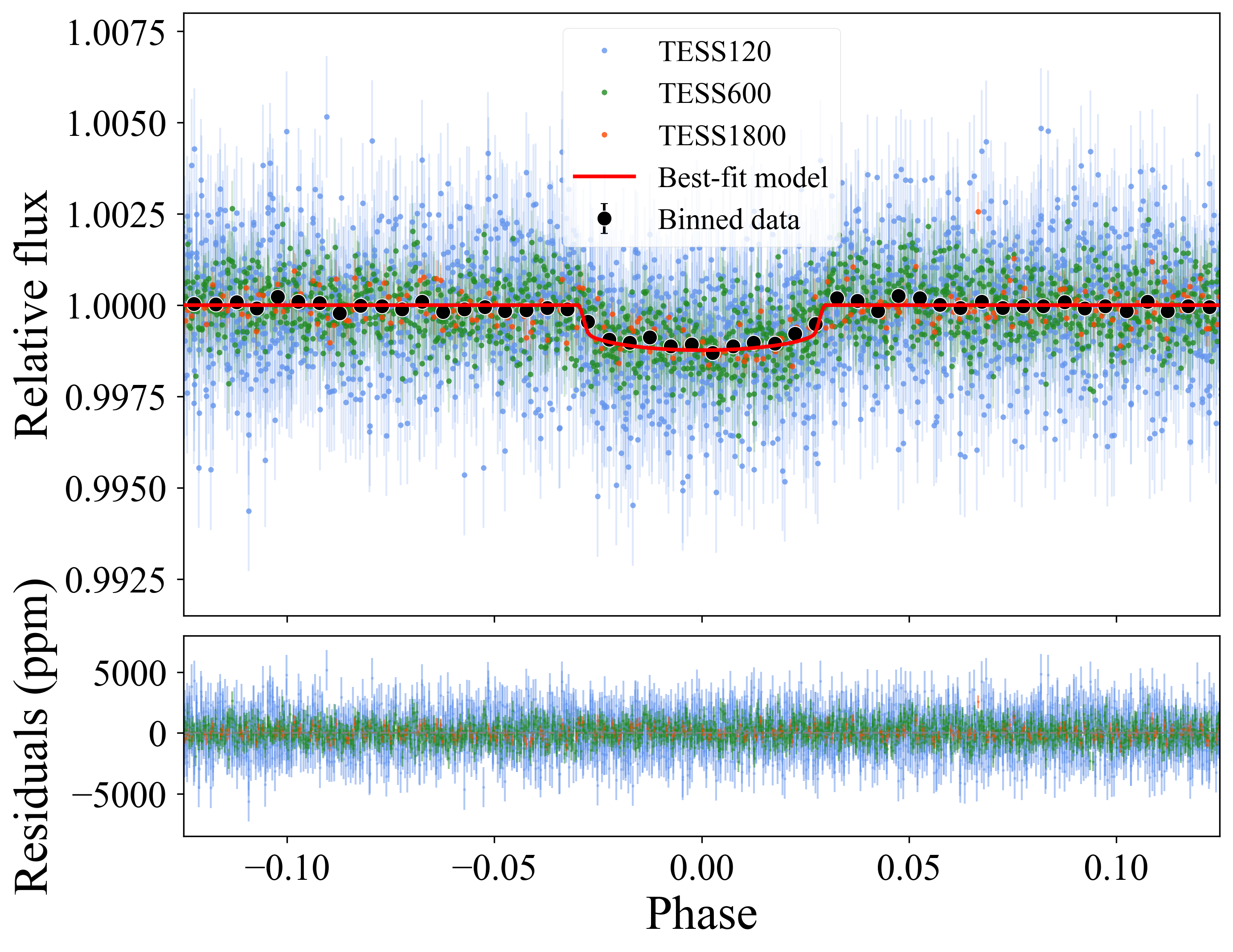}
\caption{Phase-folded unbinned TESS light curve for TOI-5646. The points are colored according to the three different exposure times used by TESS. The solid red line shows the model fit to the light curve. The error bars include both the data uncertainty and the jitter derived from the analysis. The larger black-filled circles represent the light curves binned in phase. The residuals are also shown in the bottom panel. } 
\label{fig:TESS_model}
\end{figure}

We jointly modelled the TESS photometry, ground-based follow-up photometry, and HARPS-N RV data, employing the consolidated procedure detailed by \citet{naponiello2022,naponiello2023}. In short, we modelled the transit and RV data using the \texttt{juliet}\footnote{\texttt{\url{https://juliet.readthedocs.io/}}} Python wrapper \citep{espinoza2019}, which integrates \texttt{batman}\footnote{\url{https://github.com/lkreidberg/batman}.} \citep{kreidberg2015} and \texttt{RadVel}\footnote{\url{https://radvel.readthedocs.io}.} \citep{fulton2018}. Following our previous studies (see, e.g., \citealt{naponiello2022}), photometric systematics were accounted for with Gaussian processes (GPs) using a simple Matern kernel, which was implemented in \texttt{juliet} via the \texttt{celerite} package \citep{foreman2017}. To explore posterior distributions and evaluate Bayesian evidence, we employed a Bayesian framework based on the dynamic nested-sampling tool \texttt{dynesty} \citep{speagle2020}. We adopted standard parameterisations for the orbital and transit parameters. Following \citet{eastman2013}, we parametrised the eccentricity, $e$, and the argument of the periastron, $\omega$, as $\sqrt{e}\sin{\omega}$ and $\sqrt{e}\cos{\omega}$. The impact parameter, $b$, and the ratio of the radii of the star and the planet, $R_{\rm p}/R_{\star}$, were parametrised following the approach of \citet{espinoza2018}. The limb-darkening (LD) coefficients were parametrised following \citet{kipping2013}. The dilution factor was fixed to 1.
The eccentricity was estimated at $e=0.034_{-0.024}^{+0.044}$; considering that this result is consistent with a circular orbit at the $\lesssim2\,\sigma$ level, we give a 95\% confidence upper limit of $e<0.12$. Having fixed the eccentricity to zero, we re-ran the joint fit. We found that the difference in Bayesian evidence compared to the case with variable eccentricity is $\Delta \ln{\mathcal{Z}} \gg 5$, i.e. above the very {\it strong evidence} threshold defined by \citet{kass1995}.
The posterior estimates of the main parameters of this last fit are reported in Table~\ref{tab:planet_TOI-5646}. 
The HARPS-N RVs are shown in Fig.~\ref{fig:HARPS-N_rv_fit_phased} (phase-folded), together with the preferred global model (top panels) and its residuals (bottom panels). The TESS and MuSCAT2 light
curves, folded with the planet’s orbital period, are plotted in Figs.~\ref{fig:TESS_model} and \ref{fig:muscat} together with the best-fit transit models resulting from the global fit. 

We confirm the planetary nature of the TESS candidate TOI\,5646.01, hereafter referred to as TOI-5646\,b, with a mass of $M_{\rm p}$\,=\,$45.93^{+4.46}_{-4.41}\, M_{\oplus}$ ($>$\,5\,$\sigma$ significance), a radius of $R_{\rm p}$\,=\,$6.61\pm 0.28\, R_{\oplus}$, corresponding to a bulk density of $\rho_{\rm p}$\,=\,$0.87^{+0.15}_{-0.13}$\,g\,cm$^{-3}$ and an equilibrium temperature of $T_{\rm eq}$\,=\,$1935^{+29}_{-28}$\,K.


\begin{table}
\caption{Orbital and physical parameters for TOI-5646\,b.} %
\label{tab:planet_TOI-5646} %
\centering %
\resizebox{\hsize}{!}{
\begin{tabular}{lcc}
\hline\hline \\ [-8pt] %
~~~~~~~Parameter & Unit & Value \\ [2pt] %
\hline \\[-8pt] %
\multicolumn{1}{l}{~~\textbf{LC parameters}} \\ %
$P_{\rm orb}$         \dotfill & day & 2.4277027\,(25)                      \\ [2pt] %
$T_{\rm 0}$           \dotfill & BJD$-2459000$ & 687.3808\,(11)             \\ [2pt] %
$T_{\rm 14}$          \dotfill & hr            & $3.461 \pm 0.048$          \\ [2pt] %
$i$                   \dotfill & deg & $84.77^{+0.51}_{-0.65}$              \\ [2pt] %
$b$                   \dotfill & \dotfill & $0.457^{+0.046}_{-0.038}$       \\ [2pt] %
$R_{\rm p}/R_{\star}$ \dotfill & \dotfill & $0.03300^{+0.00059}_{-0.00058}$ \\ [2pt] %
$a/R_{\star}$         \dotfill & \dotfill & $5.004^{+0.098}_{-0.111}$       \\ [2pt] %
$q_1$,\,\textsc{tess}$^{(a)}$\dotfill & \dotfill & $0.25^{+0.13}_{-0.11}$   \\ [2pt] %
$q_2$,\,\textsc{tess}$^{(a)}$\dotfill & \dotfill & $0.51^{+0.31}_{-0.30}$   \\ [2pt] %
$q_1$,\,\textsc{MuSCAT2-$g$}$^{(a)}$\dotfill & \dotfill & $0.94^{+0.04}_{-0.08}$ \\ [2pt] %
$q_2$,\,\textsc{MuSCAT2-$g$}$^{(a)}$\dotfill & \dotfill & $0.86^{+0.10}_{-0.16}$ \\ [2pt] %
$q_1$,\,\textsc{MuSCAT2-$r$}$^{(a)}$\dotfill & \dotfill & $0.62^{+0.25}_{-0.32}$ \\ [2pt] %
$q_2$,\,\textsc{MuSCAT2-$r$}$^{(a)}$\dotfill & \dotfill & $0.46^{+0.32}_{-0.29}$ \\ [2pt] %
$q_1$,\,\textsc{MuSCAT2-$i$}$^{(a)}$\dotfill & \dotfill & $0.64^{+0.22}_{-0.21}$ \\ [2pt] %
$q_2$,\,\textsc{MuSCAT2-$i$}$^{(a)}$\dotfill & \dotfill & $0.59^{+0.25}_{-0.27}$ \\ [2pt] %
$q_1$,\,\textsc{MuSCAT2-$z$}$^{(a)}$\dotfill & \dotfill & $0.71^{+0.20}_{-0.31}$ \\ [2pt] %
$q_2$,\,\textsc{MuSCAT2-$z$}$^{(a)}$\dotfill & \dotfill & $0.63^{+0.25}_{-0.34}$ \\ [4pt] %
\multicolumn{1}{l}{~~\textbf{RV parameters}} \\ [2pt] %
$K$                   \dotfill & m\,s$^{-1}$ & $17.34^{1.55}_{1.57}$ \\ [2pt] %
$\mu_{\rm{HARPS-N}}$$^{(b)}$   \dotfill & m\,s$^{-1}$ & $-8531.93^{+1.03}_{-1.07}$ \\ [2pt]
$e^{(c)}$             \dotfill & \dotfill & $e<0.12$ \\ [4pt] %
%
\multicolumn{1}{l}{\textbf{Instrumental parameters}} \\ [2pt] %
$\sigma_{\rm{TESS}}$\,{\tiny (Sect.\,23)}$^{(d)}$ \dotfill & ppm & $3.12^{+18.44}_{-2.72}$        \\ [2pt]
$\sigma_{\rm{TESS}}$\,{\tiny (Sect.\,46)}$^{(d)}$ \dotfill & ppm & $5.87^{+20.49}_{-5.12}$        \\ [2pt]
$\sigma_{\rm{TESS}}$\,{\tiny (Sect.\,50)}$^{(d)}$ \dotfill & ppm & $5.87^{+20.49}_{-5.12}$        \\ [2pt]
$\sigma_{\rm{TESS}}$\,{\tiny (Sect.\,91)}$^{(d)}$ \dotfill & ppm & $2.73^{+17.09}_{-2.37}$        \\ [2pt]
$\sigma_{{\rm MuSCAT2}-g}$$^{(d)}$                \dotfill & ppm & $51.16^{+186.32}_{-44.65}$     \\ [2pt]
$\sigma_{{\rm MuSCAT2}-r}$$^{(d)}$                \dotfill & ppm & $524.41^{+139.52}_{-220.51}$   \\ [2pt]
$\sigma_{{\rm MuSCAT2}-i}$$^{(d)}$                \dotfill & ppm & $10.60^{+62.87}_{-9.64}$       \\ [2pt]
$\sigma_{{\rm MuSCAT2}-z}$$^{(d)}$                \dotfill & ppm & $3.93^{+66.84}_{-3.60}$        \\ [2pt]
$\sigma_{\rm{HARPS-N}}$$^{(d)}$                   \dotfill & m\,s$^{-1}$ & $1.72^{+1.23}_{-1.09}$ \\ [4pt]
\multicolumn{1}{l}{~~~~~~~~\textbf{Derived}} \\ [2pt] %
$M_{\rm p}$          \dotfill & $M_{\oplus}$ & $45.93^{+4.46}_{-4.41}$   \\ [2pt] %
$R_{\rm p}$          \dotfill & $R_{\oplus}$ & $6.61\pm 0.28$ \\ [2pt] %
$\rho_{\rm p}$       \dotfill & g\,cm$^{-3}$ & $0.87^{+0.15}_{-0.13}$    \\ [2pt] %
$g_{\rm p}$          \dotfill & m\,s$^{-2}$ & $10.28^{+1.41}_{-1.27}$     \\ [2pt] %
$a$                  \dotfill & au & $0.0427 \pm 0.0019$        \\ [2pt] %
$T_{\rm eq}$$^{(e)}$ \dotfill & K  & $1935^{+29}_{-28}$         \\ [2pt] %
$\log_{14}{\langle F \rangle}$$^{(f)}$\dotfill & cgs & $9.503^{+0.052}_{-0.051}$ \\ [2pt] %
TSM$^{(g)}$          \dotfill & \dotfill & $35.78^{+4.77}_{-4.06}$       \\ [2pt] %
\hline
\end{tabular}
}
\tablefoot{The median values of the best-fit parameters for TOI-5646\,b, along with their upper and lower $68\%$ credibility intervals as uncertainties. These values were obtained from the posterior distributions of the corresponding models. The numbers in brackets represent the uncertainties in the preceding digits.
$^{(a)}$$q_1 \equiv (u_1+u_2)^2$ and $q_2 \equiv (u_1/2)(u_1+u_2)^{-1}$, where $u_1$ and $u_2$ are the LD coefficients of the quadratic law \citep{kipping2013}.
$^{(b)}$This is the systemic RV for HARPS-N.
$^{(c)}$This is the $95\%$ confidence upper limit on the eccentricity and was determined when $\sqrt{e} \cos{\omega}$ and $\sqrt{e} \sin{\omega}$ were allowed to vary in the fit.
$^{(d)}$$\sigma_{\rm{TESS}}$ and $\sigma_{\rm{HARPS-N}}$ are jitters added in quadrature to the error bars of TESS and HARPS-N, respectively.
$^{(e)}$This represents the equilibrium temperature assuming a Bond albedo of zero and a uniform redistribution of heat to the night side.
$^{(f)}$Incoming flux per unit surface area, averaged over the orbit.
$^{(g)}$Transmission spectroscopy metric (TSM; \citealt{kempton2018}).
}
\end{table}

\subsection{TTV analysis}
To search for perturbations potentially induced by additional non-transiting planets, or by bodies undetected by TESS, we investigated the presence of transit timing variations (TTVs) following the approach described in \citet{Naponiello2026b}. We found no compelling evidence of significant timing modulations, placing an upper limit of $\sim$\,10 minutes on the TTV semi-amplitude. In particular, the difference in BIC between the linear and sinusoidal models falls slightly below the commonly adopted significance threshold of 10, while the scatter statistic, $\chi^2_{\mathrm{mod}}$ \citep{Naponiello2026b}, remains below unity. This further supports the absence of a detection.

\section{Discussion}
\label{sec:discussion}
\begin{figure}
\centering
\includegraphics[width=9.0cm]{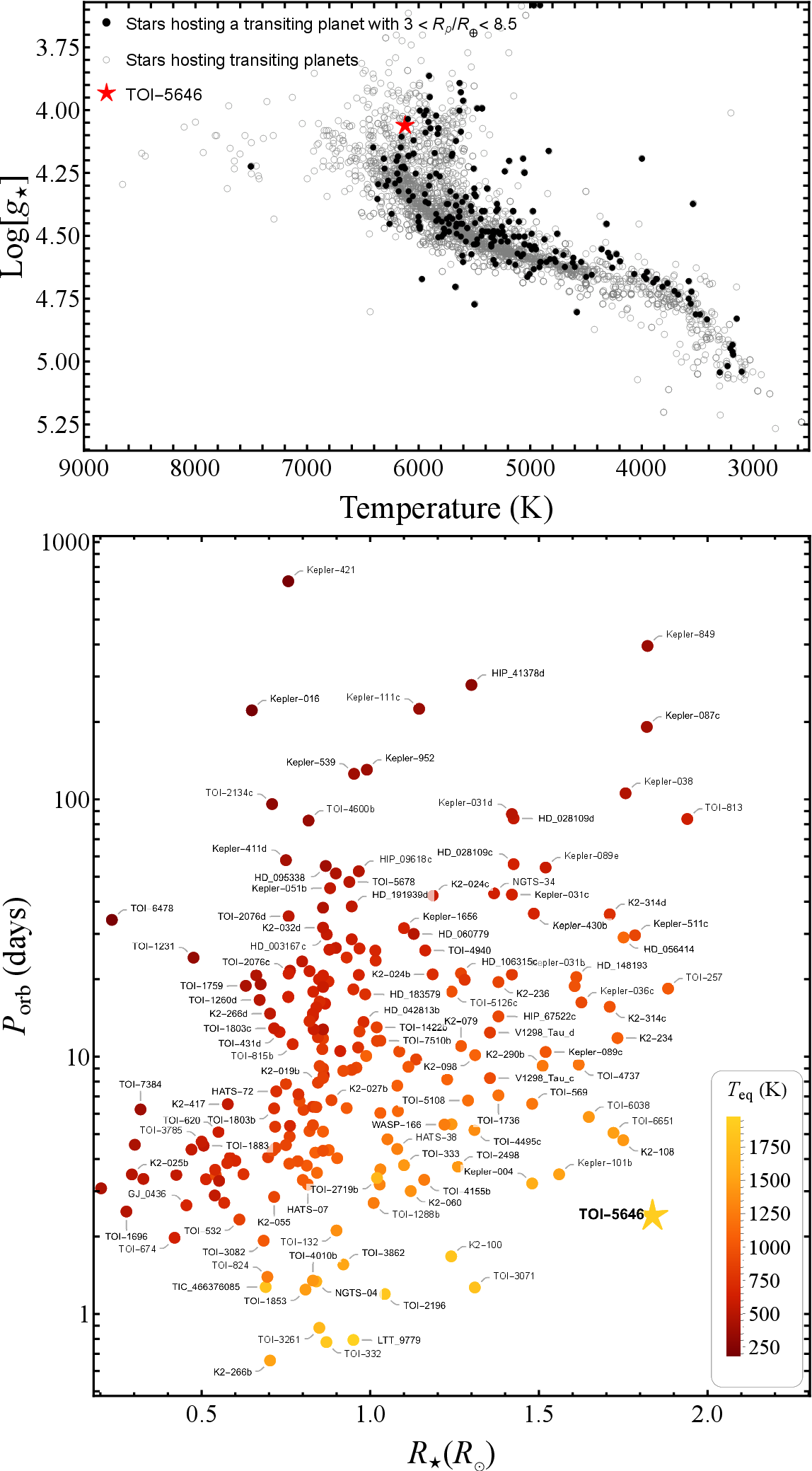}
\caption{Properties of the TOI-5646 system from a comparative
perspective. {\it Top panel}: Stellar logarithmic surface gravity vs. effective temperature for all the stars hosting at least one transiting planet (grey circles) and at least one transiting planet with a radius between 3 and 8.5 $R_{\oplus}$ (black points). {\it Bottom panel}: Orbital period of all known transiting planets with a measured equilibrium temperature, having 3\,$R_{\oplus}$\,$<$\,$R_{\rm p}$\,<\,8.5\,$R_{\oplus}$, against their parent star radii. The colour indicates the equilibrium temperature of the planets. The data are taken from {\tt TEPCat} (Transiting Extrasolar Planet Catalogue; \citealt{southworth2011}). The star marks the positions of the TOI-5646 planetary system in both panels.} 
\label{fig:diagram1}
\end{figure}

\begin{figure}
\centering
\includegraphics[width=9.0cm]{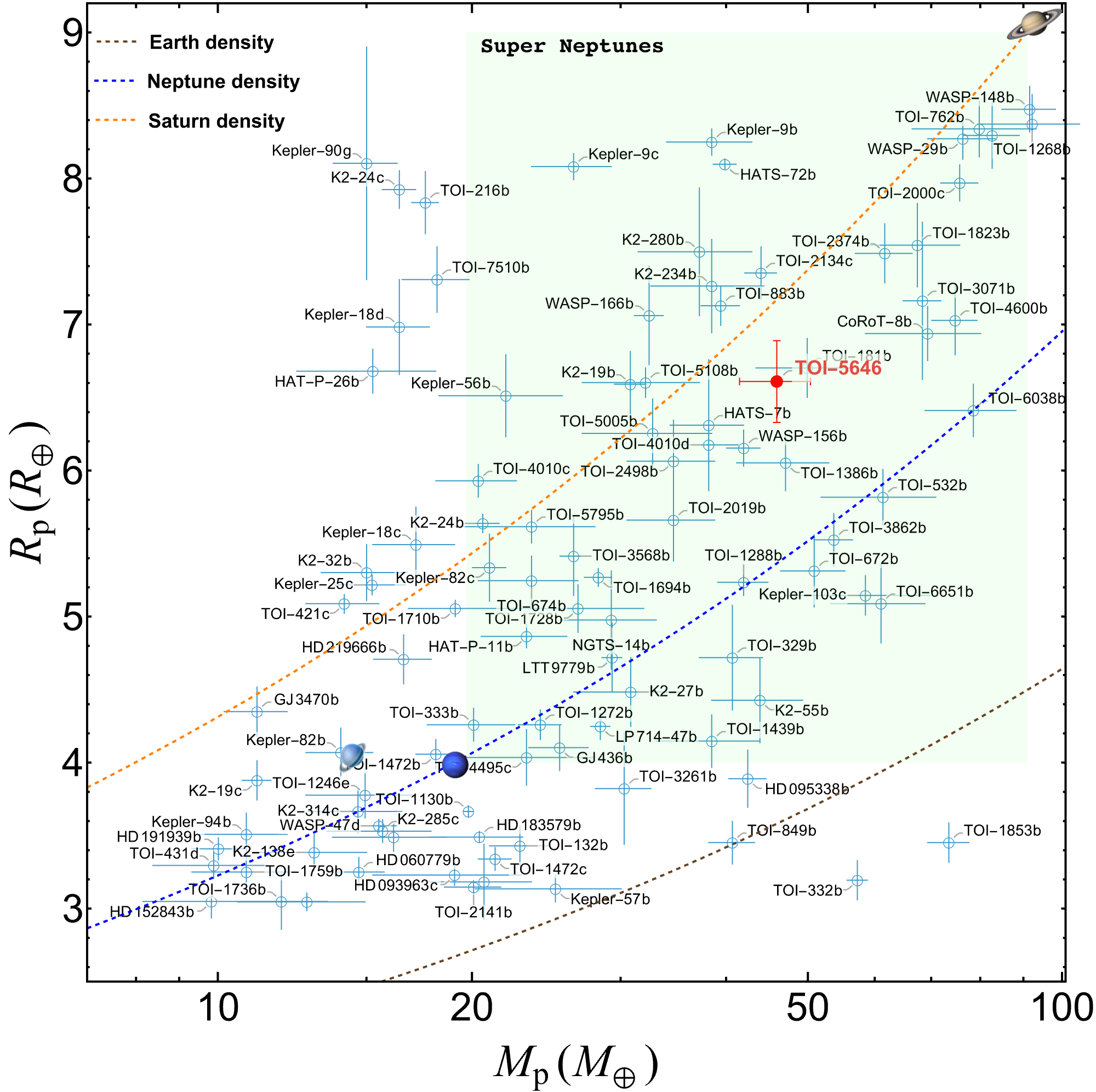}
\caption{Section of the $R_{\rm p}$ vs. $M_{\rm p}$ linear-log diagram of known transiting exoplanets with a mass between 10 and 100\,$M_{\oplus}$ (measured with an accuracy to within $20\%$) and a radius between 3 and 8.5\,$R_{\oplus}$ (measured with an accuracy to within $10\%$). The data are taken from {\tt TEPCat}. The position of TOI-5646\,b is highlighted (red point). Exoplanets inside the green zone are known exoplanets with sizes and masses between those of Neptune and Saturn. The positions of Saturn, Uranus, and Neptune are also highlighted.} 
\label{fig:RpVsMp}
\end{figure}

The top panel of Fig.~\ref{fig:diagram1} shows the $\log{g_{\star}}$ of TOI-5646 versus its $T_{\rm eff}$ compared to all other transiting planetary systems, highlighting its ongoing evolution off the main sequence. As shown in the bottom panel of Fig.~\ref{fig:diagram1}, TOI-5646 is the largest star found to host a hot planet in the Neptune desert.
The mass and size of TOI-5646\,b firmly place it in the super-Neptune category (see Fig.~\ref{fig:RpVsMp}). Its equilibrium temperature of $\sim$\,1935\,K is remarkably high, making it the second hottest planet ever discovered in the Neptune desert after LTT\,9779\,b \citep{jenkins2020}.
The relatively short orbital period of the planet and large stellar radius make TOI-5646\,b a potentially suitable target for atmospheric evaporation studies. 

\subsection{Hot Neptunes orbit metal-rich stars}
\label{sec:hotNeptunesMetal-richStars}
\begin{figure*}
\centering
\includegraphics[width=18cm]{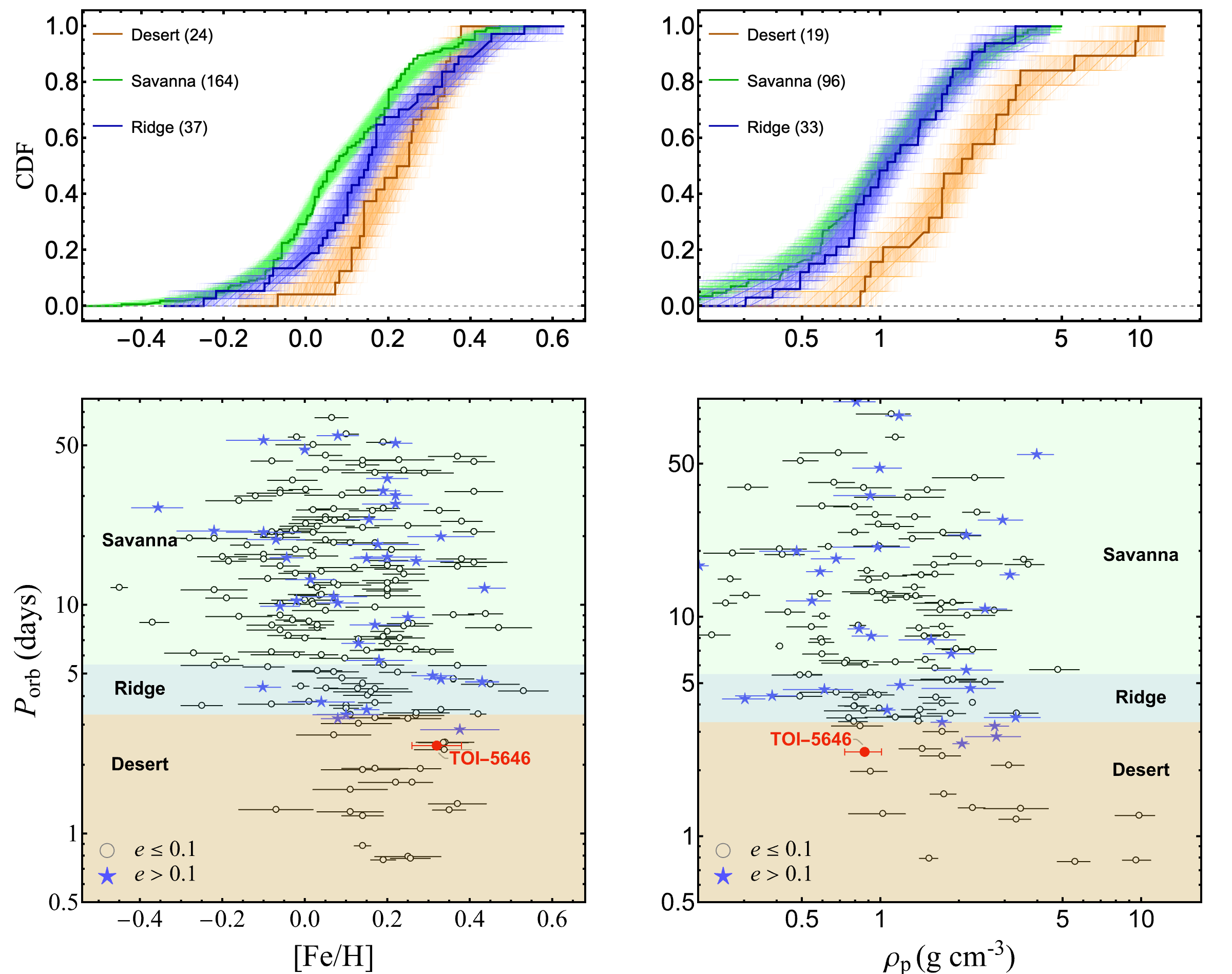}
\caption{Comparison of parent-star metallicity and planetary-density CDFs for different planet samples. {\it Bottom left panel}: Log-linear diagram of the planets' orbital period versus the corresponding parent-star metallicity with an uncertainty lower than 0.1\,dex. The circles indicate planets with an eccentricity $e \leq 0.1$, and the blue stars indicate those with $e > 0.1$. {\it Bottom right panel}: Log-log diagram of the orbital period vs. the density of known transiting exoplanets with 3\,$R_{\oplus}$\,<\,$R_{\rm p}$\,<\,8.5\,$R_{\oplus}$ and a mean density measured with an accuracy to within $30\%$. The vertical error bars are suppressed for clarity. The position of TOI-5646\,b is highlighted. The dashed-orange lines delimit the three regions recognised by \citet{castro2024a}.
{\it Top left panel}: Comparison of host-star metallicity CDFs for different Neptune-sized planet samples (3\,$R_{\oplus}$\,<\,$R_{\rm p}$\,<\,8.5\,$R_{\oplus}$) belonging to the desert (orange), ridge (blue), and savanna (green). We only considered planets orbiting stars with a metallicity known with uncertainties $<0.1$\,dex. {\it Top right panel}: Comparison of planetary density CDFs for the three aforementioned samples. We only considered planets with a mean density measured with an accuracy to within $30\%$. In the two top panels the solid curves represent the empirical CDF calculated from the data points. The light-coloured step-curves define the confidence intervals of the distributions. The numbers in brackets indicate the number of elements in each sample. The data are taken from {\tt TEPCat} in all cases.} 
\label{fig:three_zones}
\end{figure*}

As noted by several authors \citep{dong2018,vissapragada2025,doyle2025,mancini2026}, the average metallicity of host stars of Neptune-sized exoplanets located within the desert and the ridge is  higher than those found in the savanna.
This trend is clearly illustrated in the bottom left panel of Fig.~\ref{fig:three_zones} (most of the planets in the desert and in the ridge are hosted by stars with ${\rm [Fe/H]}>0$) and in the top left panel of Fig.~\ref{fig:three_zones}, where we compare the cumulative distribution functions (CDFs) of host-star metallicities in the Neptune desert, ridge, and savanna samples. For this analysis, we included all planets with measured radii between 3 and 8.5 $R_{\oplus }$ orbiting stars with known metallicities (including TOI-5646\,b), which were measured with an uncertainty lower than 0.1\,dex. The solid curves represent the empirical CDFs calculated from the nominal data points. To illustrate the impact of measurement uncertainties, the individual stellar metallicities were resampled 1000 times from their respective normal distributions (defined by the measured value and its associated error); the resulting 1000 resampled CDFs are displayed as the light step-curves, visually defining the confidence intervals of the distributions.
The null hypothesis that the Neptune-desert and Neptune-savanna datasets have the same distribution is rejected at the 5\% level based on the Kolmogorov-Smirnov test
($p=2.0\times10^{-4}$), whereas it is not rejected for the Neptune-desert and Neptune-ridge datasets ($p = 0.14$). Therefore, based on a sample of 225 Neptune-sized exoplanets, we confirm the tendency of hot Neptunes to orbit metal-rich stars. The new planet, TOI-5646\,b, which we report in this work, agrees well with this trend.

\subsection{Density -- period space}
The bottom right panel of Fig.~\ref{fig:three_zones} shows that the Neptune desert is populated by planets that are generally denser than those of the ridge and savanna, with several extreme cases. To investigate the solidity of this trend, which was noted by several authors (e.g. \citealt{castro2024,manni2025,mancini2026}), we compare in the top right panel of Fig.~\ref{fig:three_zones} the CDF related to the planetary density of the Neptune-desert sample with that of the ridge and savanna samples.
For this analysis, we considered all planets with measured radii between 3 and 8.5 $R_{\oplus }$ and with a mean density measured with an accuracy within 30\%, including TOI-5646\,b. In this case, the null hypothesis that the Neptune-desert dataset has the same distribution as the ridge and savanna datasets is rejected in both cases at the 5\% level based on the Kolmogorov-Smirnov test ($p=1.6\times10^{-3}$ and $p=8.0\times10^{-3}$, respectively). Therefore, based on a well-studied sample of 148 Neptune-sized planets, we also confirm the tendency that hot Neptunes are denser than cold ones. Although the trend appears robust, TOI-5646\,b is one of the least dense hot Neptunes. Within the uncertainties, however, its density is very similar to that of three other super Neptunes, that is, TOI-3071\,b \citep{hacker2024}, TOI-674\,b \citep{murgas2021}, and HATS-7\,b \citep{bakos2015}. 

\subsection{Internal structure and composition of TOI-5646\,b}

\begin{figure}
\centering
\includegraphics[width=\columnwidth]{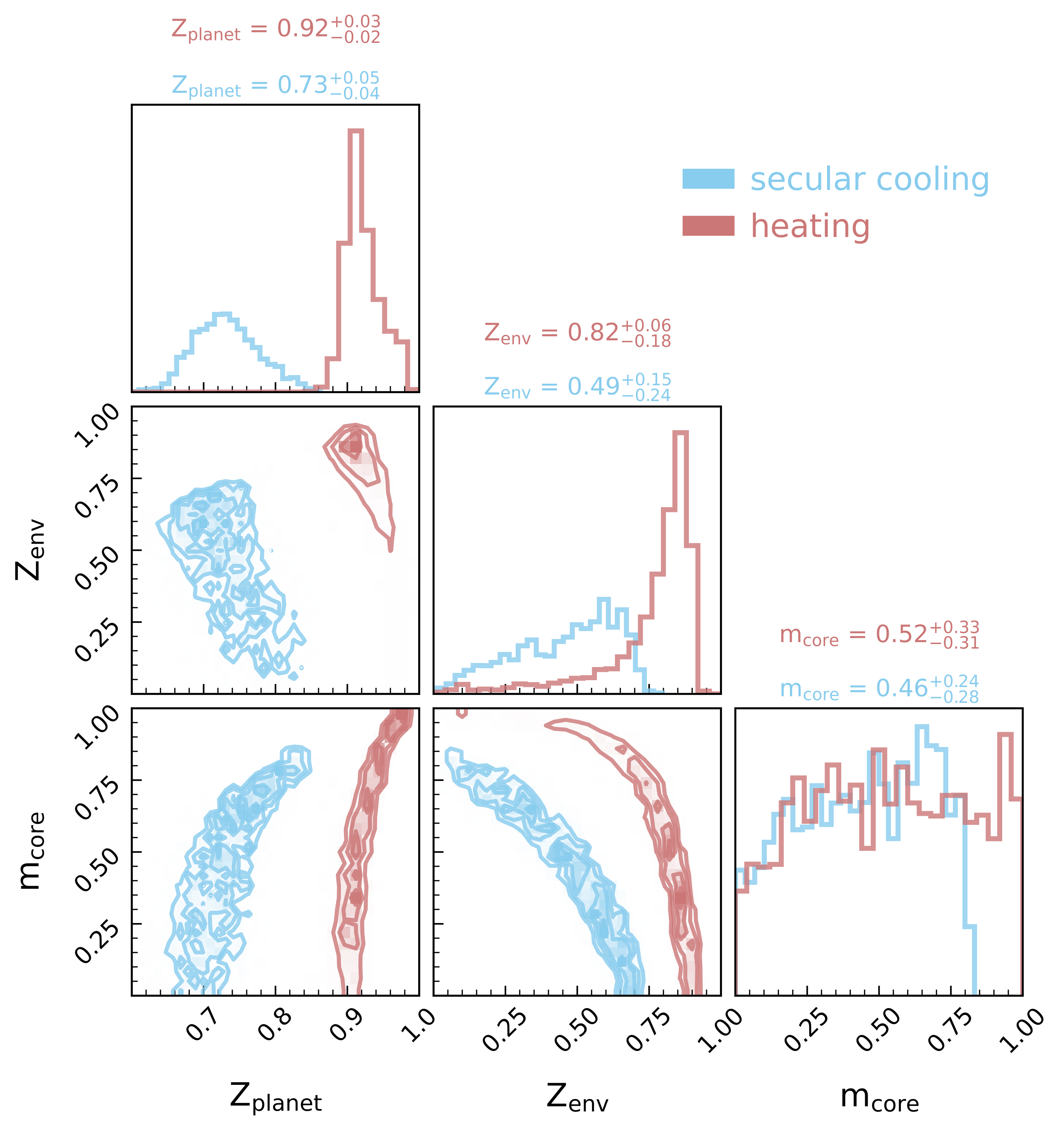}
\caption{Retrieved bulk metallicity ($Z_\mathrm{planet}$), envelope metal fraction ($Z_\mathrm{env}$), and core mass fraction ($m_\mathrm{core}$) of TOI-5646\,b given its mass and radius. The secular cooling retrieval incorporates the system age as a constraint on the intrinsic luminosity, while the heating retrieval adopts the scaling relations by \citet{thorngren2019} as a constraint on the intrinsic luminosity.}
\label{fig:corner_internal_structure}
\end{figure}

\begin{figure}
\centering
\includegraphics[width=\columnwidth]{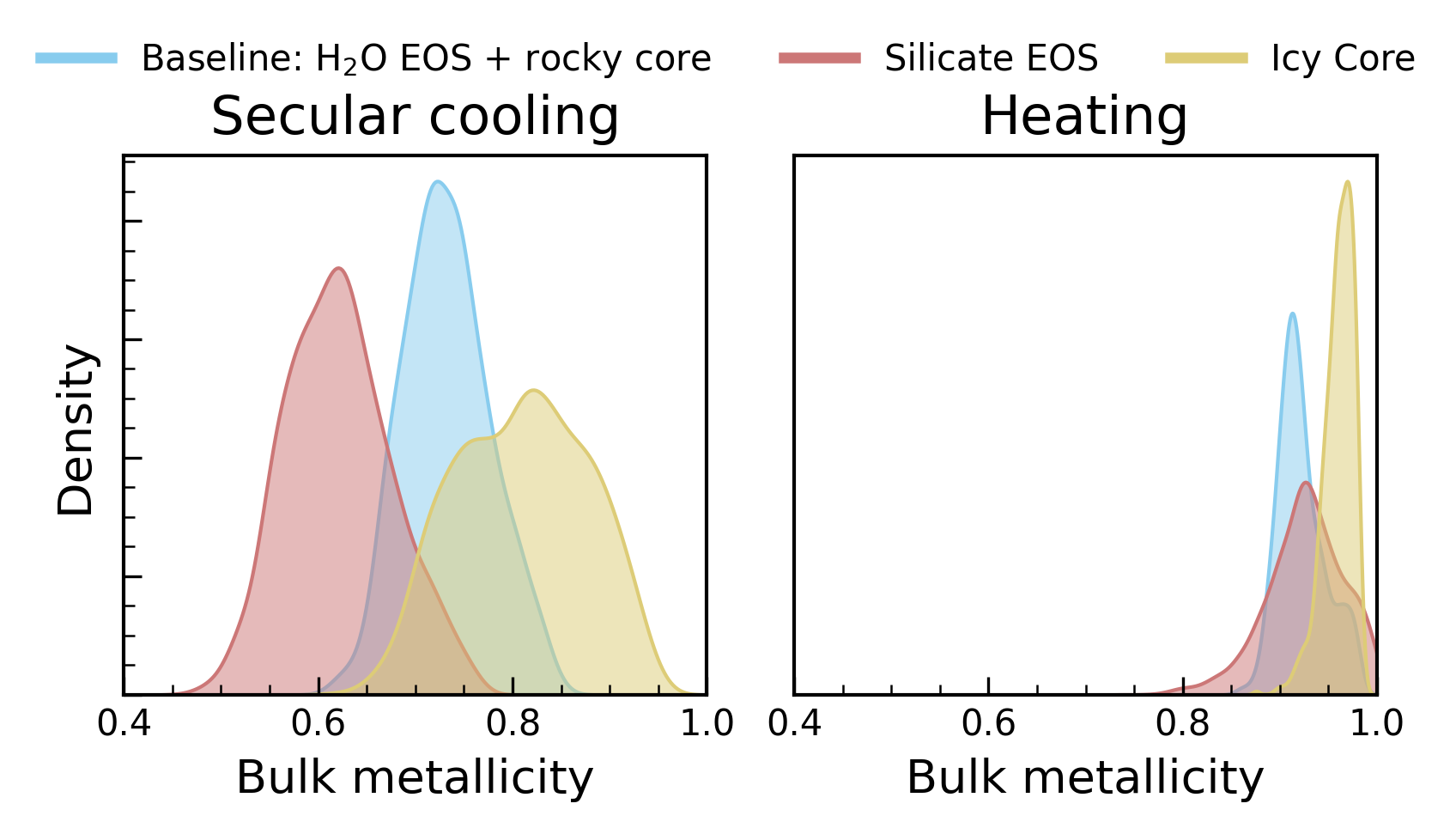}
\caption{Dependence of the retrieved bulk metallicity on the adopted equation of state (EOS) for the envelope metals and the assumed core composition. The baseline model (blue) adopts a water EOS for the envelope metals and a fully rocky core. The results obtained using a silicate EOS for the envelope metals and assuming the baseline rocky core are shown in red, and those assuming a fully icy core and the baseline water EOS for envelope metals are shown in yellow. The left panel corresponds to the secular-cooling scenario, and the right panel shows the anomalous-heating scenario.}
\label{fig:EOS_and_core_dependence}
\end{figure}

We modelled the interior structure of TOI-5646\,b with CEPAM \citep{CEPAM_Guillot}. CEPAM was originally developed to model Solar System gas giants Jupiter and Saturn, but in recent years, it has also been adapted to model the ice-giants Neptune and Uranus \citep{Ramirez_2026} and exoplanets \citep{Bloot_2023, vandijkmiguel2025}. For the super-Neptune TOI-5646\,b, we assumed a two-layer interior consisting of a rocky core, described with an analytical formula relating density to pressure \citep{Hubbard_Marley_1989}, surrounded by a homogeneous hydrogen-helium envelope enriched in metals, where water is used as a proxy for heavy elements in the equation of state (EOS). The EOSs for the envelope were taken from \citet{Chabrier2019} for hydrogen and helium, including the non-ideal mixing coefficients from \citet{HG23}, and from \citet{Mazevet2019} for water. The atmospheric boundary condition was determined using a semi-analytical non-grey atmosphere model \citep{Parmentier_2014, Parmentier_2015}.

For a given interior structure, parametrised by the core mass fraction and envelope metallicity, CEPAM computes the transit radius. This radius corresponds to the level at which the rays that graze the planetary terminator reach an optical depth of approximately unity (see Section 4.2 in \citealt{Guillot_2010} for the exact calculation). To constrain the internal structure of the planet from the observed radius, we solved the inverse problem using PyMultiNest \citep{Buchner_2014}. We adopted Gaussian priors on the planetary mass and equilibrium temperature, centred on their observed values and uncertainties, and uniform priors between 0 and 1 on the core mass fraction and envelope metal fraction. The hydrogen-to-helium ratio in the envelope was fixed to the protosolar value \citep{Lodders_2021}.

Because the interior model is static and not evolutionary, the intrinsic luminosity must be specified as an input parameter. The intrinsic luminosity is directly related to the entropy of the convective envelope and therefore affects the planetary radius.  However, for TOI-5646\,b, the appropriate intrinsic luminosity is highly uncertain because its evolutionary history is poorly constrained. The planet resides within the Neptune desert, where the formation and evolution of planets remain actively debated. Its high equilibrium temperature ($T_{\rm eq}$\,=\,1935\,K) further complicates the interpretation. Highly irradiated giant planets are frequently observed to have radii larger than predicted by standard cooling models, implying the presence of an additional heat source that delays cooling or inflates the planet \citep[e.g.][]{Guillot_2002, Demory_2011, Miller_2011}. The physical mechanism for this anomalous heating remains uncertain, and it is not known whether the same processes operate with similar efficiency for lower-mass planets such as TOI-5646\,b. In addition, the host star is relatively evolved and was substantially less luminous earlier in its history. Gas giants orbiting evolved stars have been suggested to undergo re-inflation as stellar irradiation increases \citep{Grunblatt2017}, and a similar process might affect this planet.

To account for these uncertainties, we considered two limiting evolutionary scenarios. The first assumed standard secular cooling without additional heating, and the second assumed that the planet experiences anomalous heating, either through delayed cooling during its evolution or through re-inflation at later times. For the secular cooling scenario, we adopted the intrinsic-luminosity prior derived from the age of the system and evolutionary interior models following the procedure described in Section\,2.2.1 of \citet{Bloot_2023}.

For the anomalous-heating scenario, we adopted the empirical heating efficiencies inferred for inflated hot Jupiters by \citet{Thorngren2018}. We emphasise that these relations were calibrated using planets with masses greater than $0.5M_{\rm Jup}$ and may not be directly applicable to lower-mass planets with thinner H/He envelopes such as TOI-5646\,b. Consequently, the retrieved interior compositions under this assumption should not be interpreted as measurements of the true bulk composition. Rather, given the uncertainty in the evolutionary history of the planet, they should be viewed as the outcome under an extreme high-anomalous-heating scenario, in contrast to our secular-cooling case without anomalous heating. At the system age of 2.82\,Gyr, the intrinsic luminosity associated with the residual formation heat is expected to be negligible compared to that required to explain the observed radius when anomalous heating is present. Our evolutionary models predict $L_{\rm int}$\,$\sim$\,$10^{23}$\,erg\,s$^{-1}$ from secular cooling alone, whereas the empirical heating prescription predicts intrinsic luminosities of about $10^{27}$\,erg\,s$^{-1}$ for a planet with $R_{\rm p}$\,$\approx$\,$6.6\,R_\oplus$ and $T_{\rm eq}$\,$\approx$\,1935\,K \citep{Thorngren2018}. Therefore, we assumed that the intrinsic luminosity is dominated by the anomalous-heating term and imposed a Gaussian prior based on the population-level heating efficiencies from \citet{Thorngren2018}. The mean and standard deviation of this prior were computed from the heating efficiencies given in their Eq.\,(34), with the intrinsic luminosity calculated as  $L_{\rm int} = \epsilon\sigma_{\rm B}4\pi R_{\rm p}^2 T_{\rm eq}^4$, following Equation~\,35 of \citet{Thorngren2018}, where $\epsilon$ is the heating efficiency (the fraction of incident energy converted into internal heat), $\sigma_{\rm B}$ is the Stefan-Boltzmann constant and $4\pi R_{\rm p}^2 T_{\rm eq}^4$ is the total power re-radiated by the planet at thermal equilibrium.

Fig.~\ref{fig:corner_internal_structure} shows the retrieved bulk metallicity, envelope metallicity, and core mass fraction for the two intrinsic-luminosity prescriptions. Under the heating prescription, we infer a bulk metallicity of $0.92^{+0.03}_{-0.02}$, whereas the secular-cooling model yields a lower bulk metallicity of $0.73^{+0.05}_{-0.04}$.

In both scenarios, the total heavy-element fraction is relatively well constrained, but the unknown atmospheric metallicity produces a strong degeneracy between the core mass fraction and the envelope metallicity. In the anomalous-heating case, reproducing the inferred high bulk metallicity requires either an extremely massive core or a highly metal-enriched envelope. For example, an envelope metallicity of approximately $100\times$ solar corresponds to core mass fractions between 0.8 and 0.95, whereas a more moderate core mass fraction of 0.5 requires envelope metallicities of $200-350\times$ solar. These two possibilities both appear to be relatively extreme. Envelope metallicities of $200-350\times$ solar exceed the values inferred for Neptune ($\sim$\,100$\times$ solar; \citealt{KARKOSCHKA_2011}) and Saturn ($\sim$\,10$\times$ solar; \citealt{Atreya_2022}), while core mass fractions approaching unity imply that most of the planet's mass resides in heavy elements, leaving only a thin H/He envelope.

One possible explanation for the inferred high bulk metallicity is atmospheric escape, which may have preferentially removed hydrogen and helium from the envelope over the planet lifetime, leading to a high envelope metallicity. This phenomenon might also explain the high core mass, with the planet being an almost exposed core of a once much larger planet. Alternatively, the heating efficiency could be significantly overestimated for a planet with a thinner or more metal-rich H/He envelope than inferred from the hot Jupiter population. Because higher intrinsic luminosities produce greater planetary radii, a more strongly internally heated model requires a larger heavy-element fraction to reproduce the observed radius. The secular-cooling scenario therefore results in less extreme interior compositions. In this case, an envelope metallicity of approximately $100\times$ solar is compatible with core mass fractions between 0 and 0.4,. This is more consistent with Solar System trends.

Given the high inferred heavy-element content, the retrieved bulk metallicity can also depend on the adopted metal EOS and the assumed core composition. Therefore, we investigated these systematic uncertainties by repeating the retrieval using a silicate EOS \citep{Lyon1992} for the envelope metals and an icy core composition. Fig.~\ref{fig:EOS_and_core_dependence} compares these cases with our baseline model, which adopts a water EOS for the envelope metals and a fully rocky core. 
Exchanging the baseline water EOS with a silicate EOS (red) decreases the inferred bulk metallicity in the secular-cooling case to $0.62^{+0.06}_{-0.05}$, whereas the anomalous-heating result remains essentially unchanged, with a slightly larger uncertainty ($0.93 \pm 0.04$). Exchanging the baseline rocky core with a lower-density icy core (yellow) while keeping the baseline water EOS in the envelope increases the inferred bulk metallicity to $0.81^{+0.07}_{-0.08}$ for the secular-cooling case and $0.96^{+0.01}_{-0.02}$ for the anomalous-heating case.

The efficiency of anomalous heating, the composition of the heavy elements in the envelope, and the core composition are all uncertain. Together, these different assumptions likely bracket a plausible range of bulk metallicities for TOI-5646\,b. For illustration, Fig.~\ref{fig:RpVsMp_models} presents four representative mass-radius relations corresponding to possible interior structures consistent with our analysis.

\subsection{Atmospheric evolution and formation scenario of TOI-5646\,b}
The position of TOI-5646\,b inside the Neptune desert calls for exploring which formation and atmospheric evolution paths may have led the planet to end up in this peculiar location of the parameter space. On its close-in orbit, TOI-5646\,b is exposed to extreme levels of high-energy stellar radiation (X-ray and extreme ultraviolet; together, XUV). Various models and approximations \citep[e.g.][]{Jackson2012MNRAS.422.2024J,Johnstone2021A&A...649A..96J} predict the present-day XUV stellar emission to be $\sim$\,$10^4$\,erg/s/cm$^2$. Despite the high planetary mass, the planet is likely prone to strong atmospheric escape under these conditions, which may have significantly decreased the atmospheric mass over time. If this is the case, a study of the atmospheric evolutionary path also enables us to infer some basic planetary properties at formation. 

Hydrodynamic atmospheric escape modelling applied to TOI-5646\,b \citep[using the interpolation between tabulated pre-calculated escape rates;][]{kubyshkina2018,kubyshkina2021a,kubyshkina2024} suggests present-day escape rates in the range of $3\times10^{10}-2.5\times10^{11}$\,g\,s$^{-1}$, which would already be enough to remove about 0.5\% of the planetary mass within 2.8\,Gyr. When young planets are expected to experience enhanced escape \citep[e.g.][]{kubyshkina2021b,owen2024} because the XUV emission of the young hosts is higher by an order of magnitude than that of billions of years old stars and because the atmosphere of young planets is inflated \citep[e.g.][]{lopez2012}, we expect that TOI-5646\,b lost a significant part of its atmosphere over time.

To assess the possible atmospheric escape histories of TOI-5646\,b and connect the present-day and primordial planetary parameters, we employed atmospheric evolution models based on a MESA \citep[Modules for Experiments in Stellar Astrophysics;][]{Paxton2013ApJS..208....4P,Paxton2018ApJS..234...34P} framework, accounting for planetary thermal evolution and atmospheric loss \citep{kubyshkina2021a,kubyshkina2022}. The standard stellar evolution code employed in our modelling scheme \citep[Mors;][]{Johnstone2021A&A...649A..96J} is not applicable for stellar masses above $1.2\,M_\odot$, and we therefore we extracted the stellar parameters (bolometric luminosity, effective temperature, and radius) from PARSEC models \citep[v2.0;][]{Nguyen2022A&A...665A.126N}. These parameters were then used to extract the evolution of the stellar XUV emission from scaling relations describing the age dependence of $L_{\rm X}/L_{\rm bol}$ for 1.3\,--\,1.6\,$M_{\odot}$ (0.29\,$\leq$\,$(B-V)_0$\,$<$\,0.45) stars \citep{Jackson2012MNRAS.422.2024J,McDonald2019ApJ...876...22M} and the relation of X-ray and extreme ultraviolet flux \citep{Johnstone2021A&A...649A..96J}. This information defines the planetary equilibrium temperature and XUV irradiation at the planetary orbit. 

Using the model described above, we ran a grid of 100 forward evolution models starting with initial total planetary masses ranging between 45.7\,$M_{\oplus}$ and 55.4\,$M_{\oplus}$ (5 cases) and initial atmospheric mass fractions between 0.1 and 0.5 (20 cases for each initial mass). For all models, we assumed that the planet did not migrate following protoplanetary disk dispersal set at an age of 10\,Myr. The present-day parameters of TOI-5646\,b can be reproduced by models with initial planetary masses below 54\,$M_{\oplus}$ and initial atmospheric mass fractions in the range $22-32\%$. For these models, the planet loses $6-16\%$ of its initial mass through atmospheric escape. However, the majority of these models predicted that at the current age, the planet not only still experiences strong atmospheric escape, but also contracts at a relatively high rate ($\sim$\,$0.25-0.4\,R_{\oplus}$ per Gyr).

According to planet formation models, the derived primordial parameters place TOI-5646\,b within the rare group of massive planets at short orbital separations that accreted less than 50\% of their mass as H-He \citep[see e.g. Fig.\,1 of][]{Emsenhuber2025A&A...701A..64E}. These planets are thought to be rare because they have to follow a peculiar formation scenario. Such massive planets typically form around a massive core beyond the water-ice line, which would make runaway gas accretion, unavoidable that would turn them into Jupiter-like giant planets,  \citep[e.g.][]{Emsenhuber2021A&A...656A..70E}. To explain the relatively low initial atmospheric mass fractions predicted by our models, we therefore have to assume one of the following scenarios: $i$) an early protoplanetary dispersion of the gas disk or a formation in the gas-poor part of the disk, which would have prevented the planet from reaching Jupiter-like masses; $ii$) a collision with another planet that would have stripped part of the envelope or triggered an earlier inward migration; $iii$) a later tidal destruction of the fully formed hot Jupiter \citep{Hallatt2026ApJ...997..139H}; or $iv$) atmospheric erosion through Roche-lobe overflow during an earlier high-eccentricity migration phase \citep{Yu2024ApJ...972..159Y}.

The results presented here depend somewhat on the sophistication of the internal structure modelling, for example, on whether the model accounts for the dependence of atmospheric opacities on metallicity \citep{Siebenaler2026MNRAS.546f2205S}, compositional gradients \citep[e.g.][]{Eberlein2025A&A...703A..72E}, or bloating luminosity \citep[e.g.][]{Emsenhuber2021A&A...656A..70E}. However, since accounting for these phenomena would lead to larger radii for a given atmospheric mass fraction, their inclusion would likely lead to an even lower initial atmospheric mass fraction, making the formation of TOI-5646\,b even more peculiar.

\section{Conclusions}
\label{sec:conclusions}
As part of the HONEI programme, we observed a selection of TESS transiting-exoplanet candidates, primarily hot and warm Neptune-sized planets, using high-resolution spectrographs. 
These exoplanets vary widely in their physical characteristics, which suggests heterogeneous internal structures and atmospheres. Our goal is to confirm these candidates as planets and precisely determine their orbital and physical characteristics. By expanding the population of well-characterised Neptune-sized planets, our aim is to understand their formation and migration history better.

We reported here the discovery and characterisation of TOI-5646\,b, a highly irradiated super-Neptune ($M_{\rm p}$\,=\,$45.93^{+4.46}_{-4.41}$\,$M_{\oplus}$, $R_{\rm p}=6.61\pm 0.28$\,$R_{\oplus}$) orbiting a metal-rich ([Fe/H]\,=\,+0.32\,$\pm$\,0.06) evolved ($M_{\star}$\,=\,$1.418^{+0.068}_{-0.079}$\,$M_{\sun}$, $R_{\star}$\,=\,$1.837^{+0.075}_{-0.066}$\,$R_{\sun}$) star on a short-period orbit ($\sim$\,2.4\,days). We classified the parent star as F8\,IV-V (Table~\ref{tab:star_TOI-5646}) because it is probably about to leave the main sequence to become a subgiant. 
The relatively evolved nature of TOI-5646 ($\log{g_{\star}} = 4.060_{-0.040}^{+0.034}$; see the top panel of Fig.~\ref{fig:diagram1}) may have implications for the long-term evolution of the planetary system. Because of the close proximity of the planet to its host star ($a/R_{\star} \approx 5$), tidal interactions are expected to have played an important role in the system evolution over its lifetime. The observed orbital eccentricity, which is compatible with zero ($e < 0.12$ at the 95\% confidence level), is consistent with tidal circularisation, although the corresponding timescale depends on the adopted tidal dissipation efficiency \citep{lanza2013,lanza2022,lanza2024}. Likewise, the short orbital period suggests that TOI-5646\,b is tidally synchronised.
As the host star evolves off the main sequence, its radius and luminosity increase, exposing the planet to progressively stronger irradiation than it experienced during most of its main-sequence lifetime. Although our atmospheric evolution models indicate that the bulk of the atmospheric escape was driven by intense XUV emission during the earlier stages of stellar evolution, the current intense irradiation can still contribute to the ongoing atmospheric evolution. In addition, the stellar expansion enhances the efficiency of tidal dissipation within the star, implying that orbital decay is expected to become increasingly stronger as the star continues its evolution towards later evolutionary stages. TOI-5646 therefore provides an interesting snapshot of a close-in super-Neptune-mass planet orbiting a host star at the onset of post-main-sequence evolution.

After LTT 9779\,b \citep{jenkins2020}, TOI-5646\,b is the hottest planet ever discovered in the Neptune desert and therefore represents a significant addition to the small population of survivors within the Neptune desert. 
Internal-structure modelling indicates that TOI-5646\,b is a highly metal-enriched planet, with an inferred bulk metallicity between $0.74\pm0.04$ and $0.92_{-0.02}^{+0.03}$, depending on the assumed efficiency of the radius inflation mechanisms. While the higher estimate implies an extreme composition, a more physically plausible scenario suggests an envelope metallicity of approximately $100\times$ solar and a core mass fraction of $0-0.4$.

Atmospheric evolution modelling computed by accounting for realistic mass-loss rates suggests that the planet might have endured significant atmospheric escape that shrank its radius over time. 
To reproduce the currently observed system parameters, the planet started off with a primordial mass below 54\,$M_{\oplus}$ and an initial atmospheric mass fraction in the range of $22-32\%$, with $6-16\%$ of the original mass having been lost through escape. These results, coupled with the position of the planet in the Neptune desert, indicate that the planet had a peculiar formation pathway, probably characterised by an early protoplanetary disc dispersal, a collision with another planet, or a later tidal destruction event.

TOI-5646\,b stands out as an interesting target for future atmospheric characterisation via transmission spectroscopy with the James Webb Space Telescope. Measurement of its atmospheric composition, especially its C/O ratio \citep{ashtari2026}, and its mass-loss rate will provide unprecedented insights into the survival mechanisms of intermediate-mass planets and the terminal evolution of planetary systems around stars leaving the main sequence.

\begin{acknowledgements}
This work is based on observations made with the Italian Telescopio Nazionale Galileo (TNG) operated by the Fundaci\'{o}n Galileo Galilei (FGG) of the Istituto Nazionale di Astrofisica (INAF) at the Observatorio del Roque de los Muchachos (La Palma, Canary Islands, Spain); programme A52TAC$\_$38 (PI: F. Manni). 
We acknowledge the Italian center for Astronomical Archives (IA2, \url{https://www.ia2.inaf.it}), part of the Italian National Institute for Astrophysics (INAF), for providing technical assistance, services and supporting activities of the GAPS collaboration.
This work includes data collected with the TESS mission, obtained from the MAST data archive at the Space Telescope Science Institute (STScI). Funding for the TESS mission is provided by the NASA Explorer Program. STScI is operated by the Association of Universities for Research in Astronomy, Inc., under the NASA contract NAS 5-26555. 
The authors acknowledge the use of public TESS data from pipelines at the TESS Science Office and at the TESS Science Processing Operations Center. Resources supporting this work were provided by the NASA High-End Computing Program through the NASA Advanced Supercomputing Division at Ames Research Center for the production of the SPOC data products. 
This work includes data collected with the MuSCAT2 instrument, developed by ABC, at Telescopio Carlos S\'{a}nchez operated on the island of Tenerife by the IAC in the Spanish Observatorio del Teide.
This work is partly supported by JSPS KAKENHI Grant Numbers JP24H00017, JP25K24620, JP26H01402, JP24K00689 and JP26K00755.
This research has used the Exoplanet Follow-up Observation Program (ExoFOP; DOI: 10.26134/ExoFOP5) website, which is operated by Caltech, under contract with the National Aeronautics and Space Administration under the Exoplanet Exploration Program.
This research uses the NASA Exoplanet Archive, operated by the California Institute of Technology, under contract with the National Aeronautics and Space Administration under the Exoplanet Exploration Program.
This work uses \texttt{TESS-cont} (\url{https://github.com/castro-gzlz/TESS-cont}), which also made use of \texttt{tpfplotter} \citep{aller2020} and \texttt{TESS-PRF} \citep{bell2022}.
This publication makes use of The Data \& Analysis Center for Exoplanets (DACE), which is a facility based at the University of Geneva (CH) dedicated to extrasolar planets data visualisation, exchange and analysis. 
We acknowledge financial support from the Agencia Estatal de Investigaci\'on of the Ministerio de Ciencia e Innovaci\'on
MCIN/AEI/10.13039/501100011033 and the ERDF ``A way of making Europe'' through projects PID2021-125627OB-C32 and PID2024-158486OB-C32. This work is supported by the European Union (ERC AdvG SPEAR, GA\,101200674).
L.N. acknowledges financial contribution from the INAF Large Grant 2023 ``EXODEMO''.
E.A.v.D. and Y.M. acknowledge support from the European Research Council (ERC) under the European Union's Horizon 2020 research and innovation programme (grant agreement no. 101088557, N-GINE). 
M.L.M. is supported by their individual research time under NASA contracts NAS5-26555 and NAS5-03127 to the Associated Universities for Research in Astronomy for the operation of the Hubble Space Telescope and the James Webb Telescope Science Operations Centers at STScI.
F.M. acknowledges the financial support from the Agencia Estatal de
Investigaci\'{o}n del Ministerio de Ciencia, Innovaci\'{o}n y
Universidades (MCIU/AEI) through grant PID2023-152906NA-I00.
L.M. thanks the Turin Astrophysical Observatory for the kind hospitality. 

\end{acknowledgements}

\bibliographystyle{aa} 
\bibliography{bib}

\begin{appendix}

\onecolumn

\section{Additional table}
Table~\ref{tab:RV_TOI-5646} contains the RV measurements of TOI-5646, which were obtained with HARPS-N (this work) and the corresponding activity indices, including H$\alpha$ and the $\log{R^{\prime}_{\rm HK}}$ index. FWHM, and BIS are the full width at half maximum and bisector span of the cross-correlation function, respectively. 

\begin{table}[!ht]
\centering %
\caption{TOI-5646 HARPS-N RV data points and activity indices.}
\label{tab:RV_TOI-5646}
\begin{tabular}{crlccccc}
\hline %
\hline  \\[-8pt]
BJD$_{\rm TDB}$ & $T_{\rm exp}$ & ~~~~~~~~~~~~RV & S/N & FWHM & BIS & H$\alpha$ & $\log{R^{\prime}_{\rm HK}}$  \\
$-2460000$ & (s)~ & ~~~~~~~~~(m\,s$^{-1}$) & & (m\,s$^{-1}$) &  (km\,s$^{-1}$)\\ [2 pt]
\hline  \\[-6pt] %
1055.69148 & 900  & $ -8518.9 \pm ~\,4.2 $ & 30.5 & $ 7761.44 \pm 8.45 $ & $ 0.04604 \pm 0.0085 $ & $ 0.1020 \pm 0.0013 $ & $ -4.963 \pm 0.013  $ \\
1064.77867 & 900  & $ -8545.1 \pm 12.9 $ & 12.6 & $ 7778.30 \pm 25.9 $ & $ 0.04825 \pm 0.0260 $ & $ 0.0993 \pm 0.0027 $ & $ -4.911 \pm 0.036  $ \\
1072.70419 & 900  & $ -8513.9 \pm ~\, 4.1 $ & 31.4 & $ 7764.35 \pm 8.27 $ & $ 0.04732 \pm 0.0083 $ & $ 0.0958 \pm 0.0012 $ & $ -5.063 \pm 0.014  $ \\
1080.65500 & 900  & $ -8524.1 \pm ~\, 4.9 $ & 26.7 & $ 7771.76 \pm 9.70 $ & $ 0.05251 \pm 0.0097 $ & $ 0.1020 \pm 0.0016 $ & $ -5.038 \pm 0.017  $ \\
1081.67581 & 900  & $ -8549.3 \pm ~\, 3.4 $ & 35.8 & $ 7757.55 \pm 6.83 $ & $ 0.05962 \pm 0.0068 $ & $ 0.0979 \pm 0.0011 $ & $ -5.016 \pm 0.011  $ \\
1082.67137 & 900  & $ -8509.8 \pm ~\, 3.9 $ & 32.6 & $ 7755.05 \pm 7.75 $ & $ 0.04833 \pm 0.0078 $ & $ 0.0993 \pm 0.0012 $ & $ -5.091 \pm 0.014  $ \\
1083.68847 & 900  & $ -8549.5 \pm ~\, 3.2 $ & 36.2 & $ 7760.33 \pm 6.42 $ & $ 0.04564 \pm 0.0064 $ & $ 0.0963 \pm 0.0013 $ & $ -5.016 \pm 0.010  $ \\
1087.68873 & 1200 & $ -8518.8 \pm ~\, 7.4 $ & 20.0 & $ 7765.23 \pm 14.8 $ & $ 0.05087 \pm 0.0150 $ & $ 0.0973 \pm 0.0016 $ & $ -4.988 \pm 0.027  $ \\
1089.68328 & 900  & $ -8525.7 \pm ~\, 4.3 $ & 30.4 & $ 7768.82 \pm 8.61 $ & $ 0.04338 \pm 0.0086 $ & $ 0.0987 \pm 0.0012 $ & $ -4.986 \pm 0.013  $ \\
1090.72750 & 900  & $ -8533.7 \pm ~\, 4.0 $ & 31.4 & $ 7771.83 \pm 7.92 $ & $ 0.02589 \pm 0.0079 $ & $ 0.0934 \pm 0.0013 $ & $ -5.086 \pm 0.015  $ \\
1091.70290 & 900  & $ -8533.3 \pm ~\, 3.2 $ & 37.3 & $ 7756.17 \pm 6.48 $ & $ 0.04983 \pm 0.0065 $ & $ 0.0963 \pm 0.0011 $ & $ -5.052 \pm 0.011  $ \\
1092.73190 & 900  & $ -8519.0 \pm ~\, 3.6 $ & 34.8 & $ 7761.26 \pm 7.28 $ & $ 0.04989 \pm 0.0073 $ & $ 0.0954 \pm 0.0011 $ & $ -5.031 \pm 0.012  $ \\
1099.56445 & 900  & $ -8521.2 \pm ~\, 6.6 $ & 22.1 & $ 7744.91 \pm 13.3 $ & $ 0.02658 \pm 0.0130 $ & $ 0.0995 \pm 0.0014 $ & $ -4.977 \pm 0.022  $ \\
1107.67984 & 900  & $ -8532.6 \pm ~\, 4.7 $ & 27.5 & $ 7754.57 \pm 9.37 $ & $ 0.03759 \pm 0.0094 $ & $ 0.0959 \pm 0.0015 $ & $ -5.065 \pm 0.017  $ \\
1109.64541 & 900  & $ -8519.7 \pm ~\, 4.5 $ & 28.4 & $ 7753.96 \pm 9.07 $ & $ 0.04007 \pm 0.0091 $ & $ 0.0956 \pm 0.0014 $ & $ -4.899 \pm 0.012  $ \\
1128.60807 & 1200 & $ -8517.9 \pm ~\, 5.5 $ & 25.8 & $ 7753.89 \pm 11.0 $ & $ 0.03195 \pm 0.0110 $ & $ 0.0936 \pm 0.0013 $ & $ -5.202 \pm 0.038  $ \\
1130.66325 & 1800 & $ -8527.7 \pm ~\, 8.1 $ & 18.8 & $ 7777.26 \pm 16.2 $ & $ 0.05764 \pm 0.0160 $ & $ 0.1000 \pm 0.0017 $ & $ -4.854 \pm 0.021  $ \\
\hline %
\end{tabular}
\end{table}

\section{Additional plots}

\begin{figure}[!ht]
\begin{minipage}{9.0cm}
  \centering
  \includegraphics[width=\textwidth]{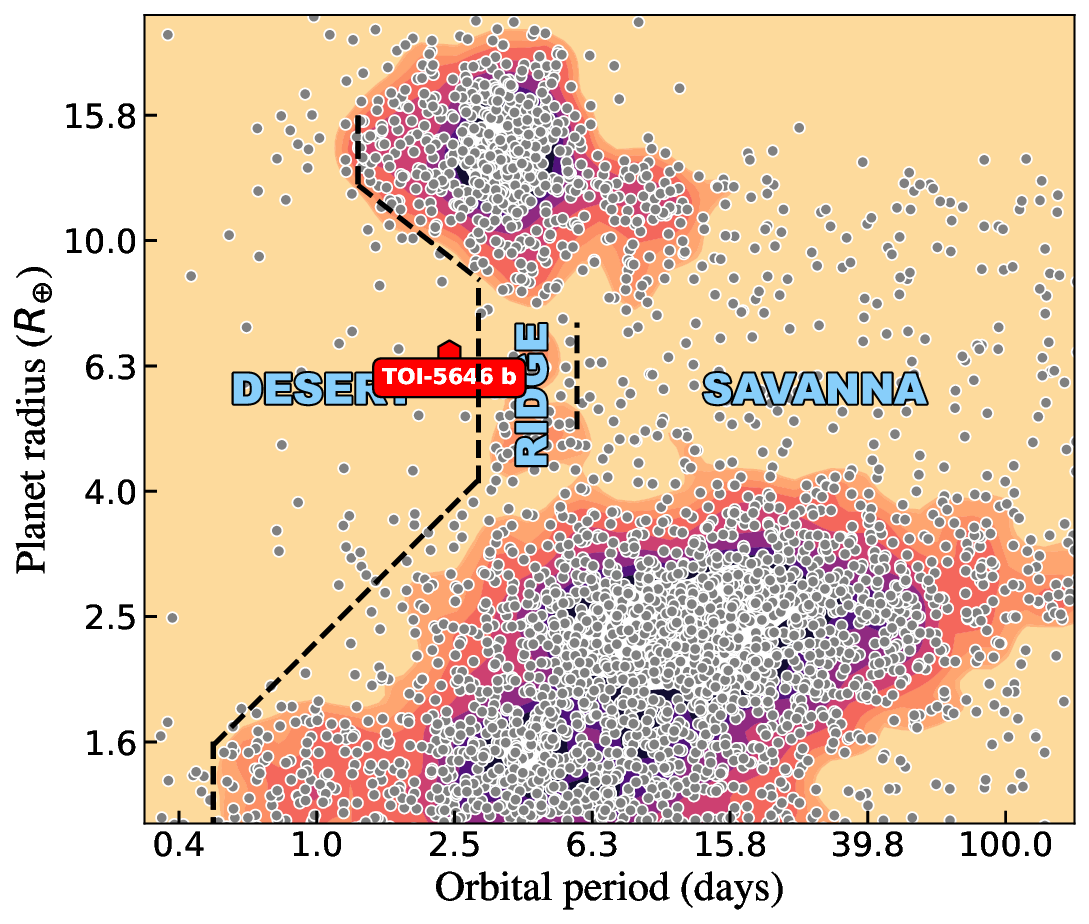}
\caption{Radius-period diagram of close-in exoplanets with mass and radius measured to an accuracy of at least $5\,\sigma$. The data were obtained from the NASA Exoplanet Archive \citep{christiansen2025} on April 21, 2026. 
Error bars have been suppressed for clarity. 
The position of TOI-5646\,b is highlighted, together with the population-based boundaries of the Neptunian desert, ridge, and savanna, as derived by \citet{castro2024}. This plot was generated with \texttt{nep-des} ({\url{github.com/castro-gzlz/nep-des}}).} 
\label{fig:nep_des}
\end{minipage}
\end{figure}
\begin{figure*}
\centering
\includegraphics[width=18.0cm]{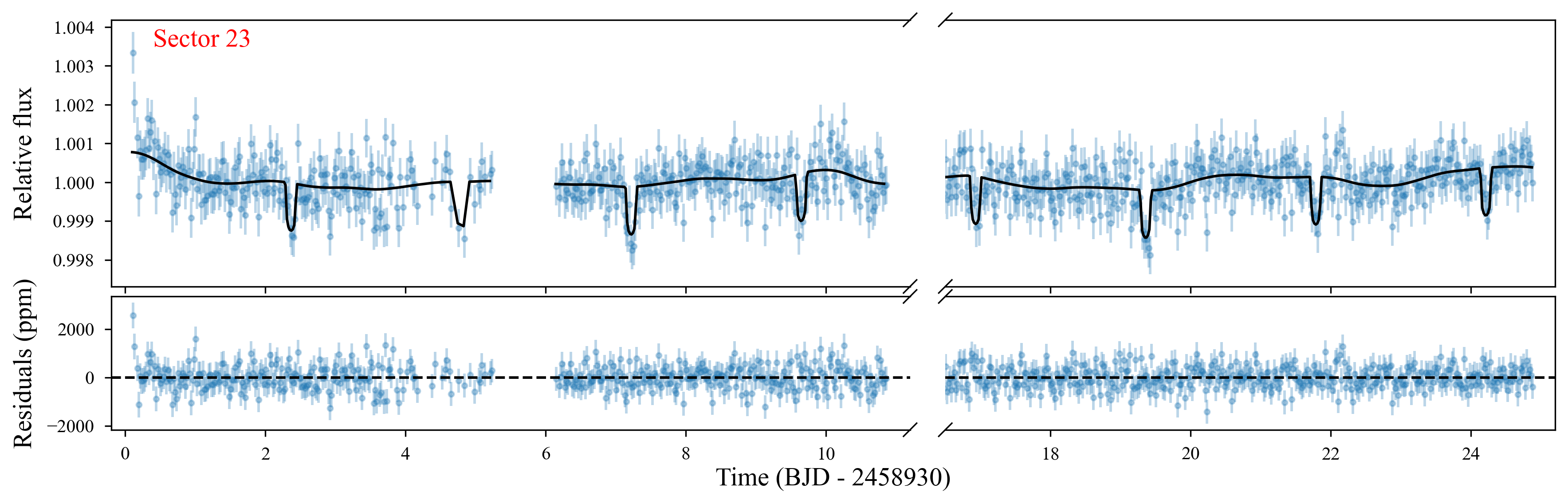}
\includegraphics[width=18.0cm]{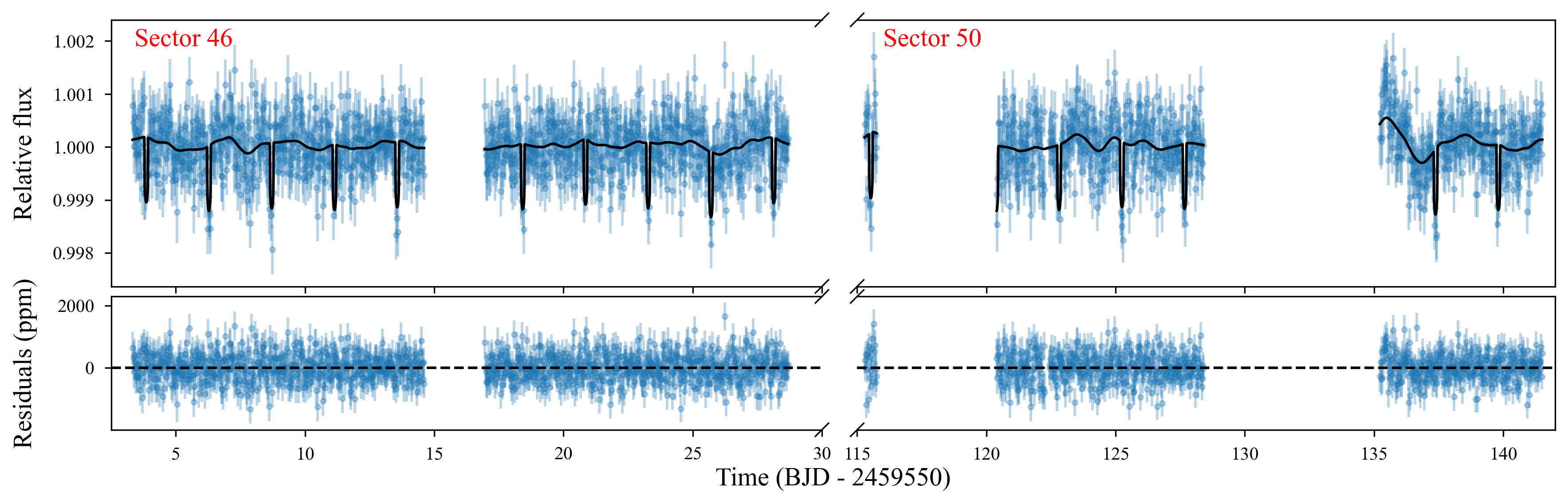}
\includegraphics[width=18.0cm]{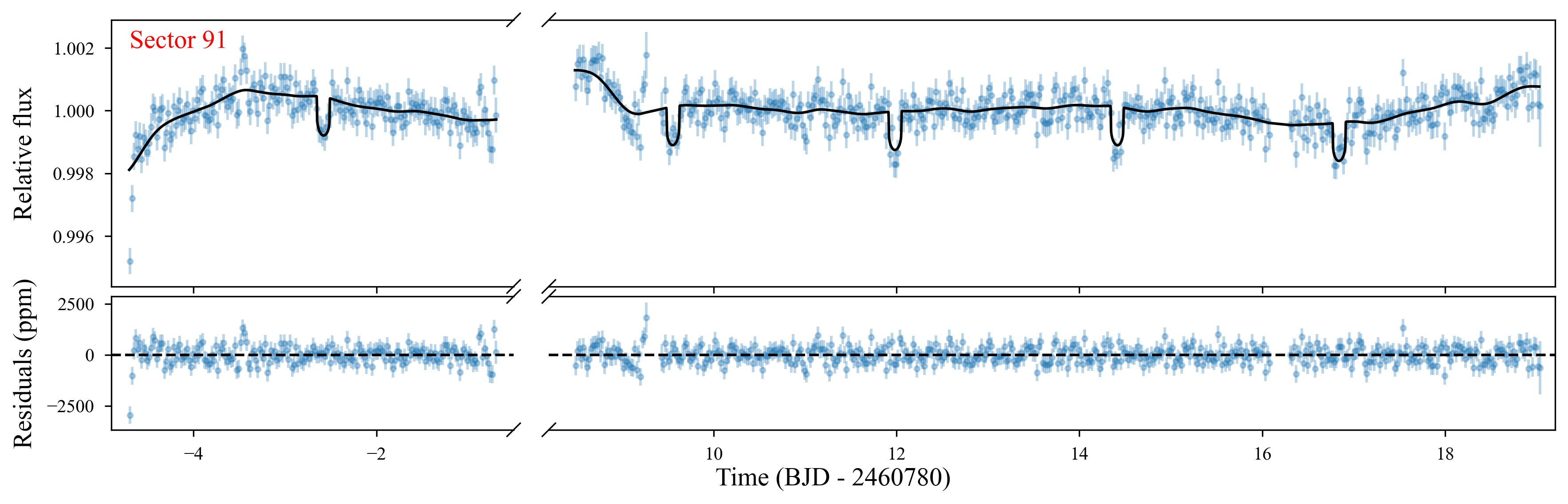}
\caption{Light curves of TOI-5646 from the PDC-SAP pipeline as collected by TESS with a 30- (top panel), 10- (middle panel) and 2-minute (bottom panel) cadence, respectively. The black line represents our best-fitting transit model (see Sect.~\ref{sec:analysis_characterisation}), whose residuals are also shown in parts per million.} 
\label{fig:TESS_lc}
\end{figure*}

\twocolumn

\begin{figure}
\centering
\includegraphics[width=9.0cm]{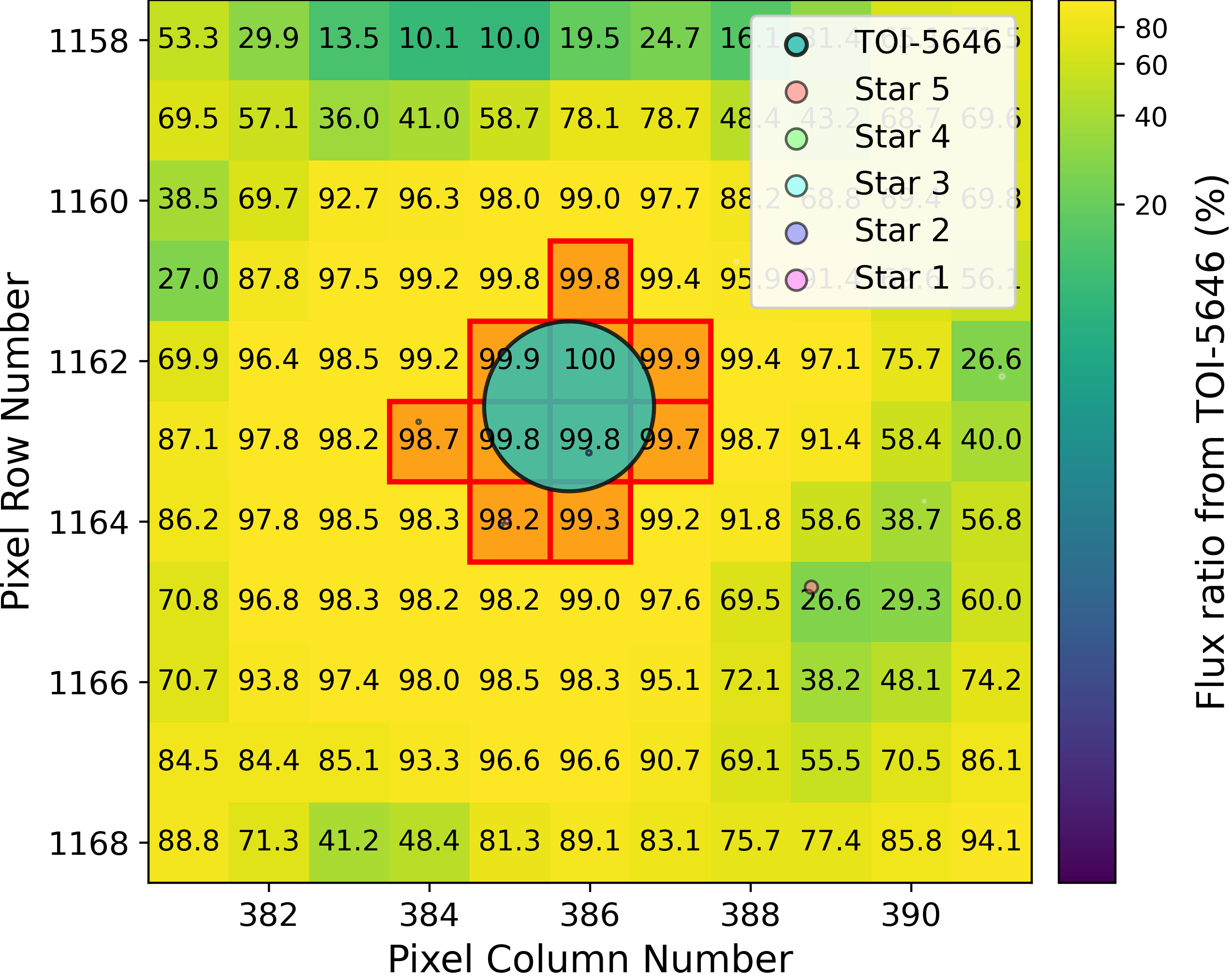}
\includegraphics[width=9.0cm]{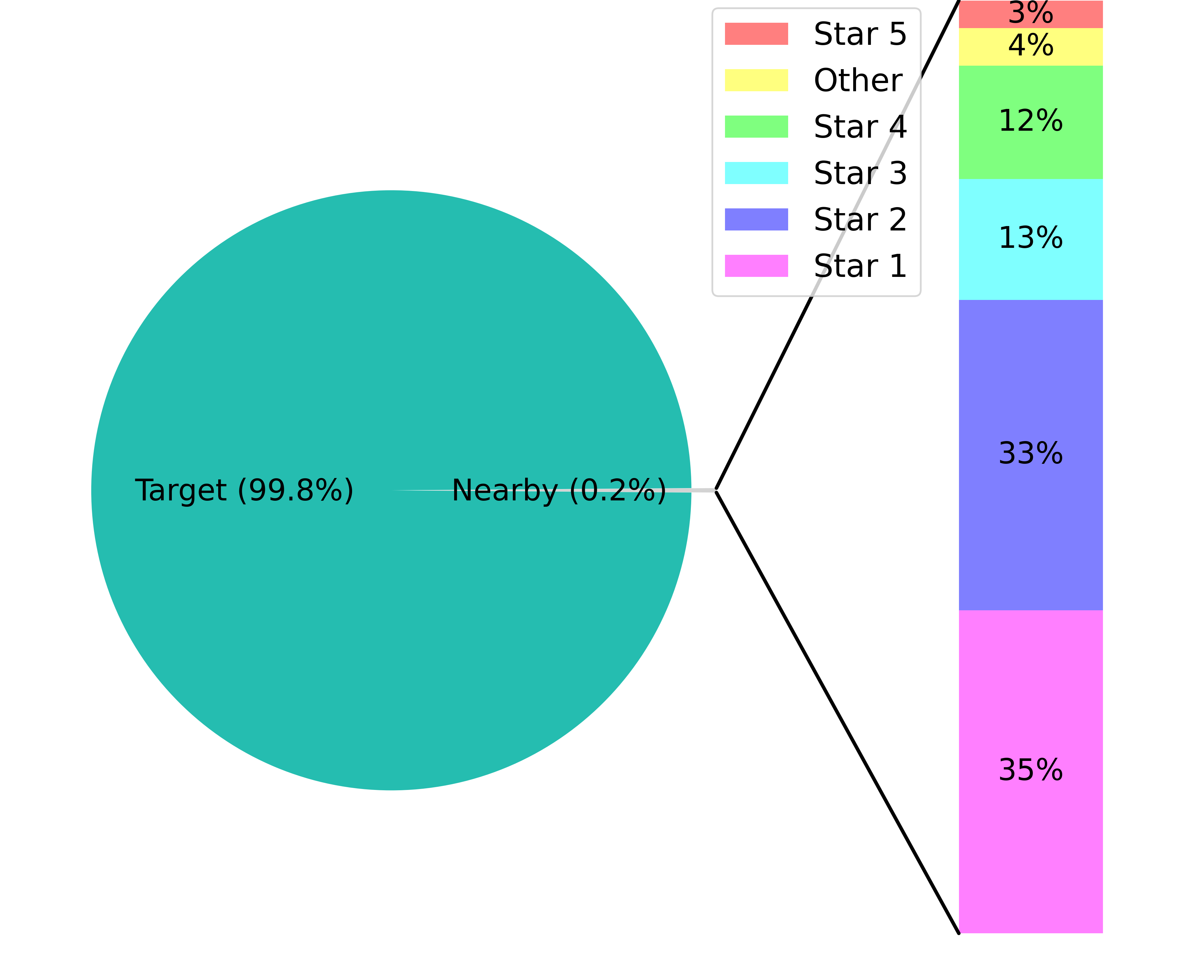}
\caption{Nearby sources contaminating the TOI-5646 photometry. {\it Top panel}: TPF-shaped heatmap with the pixel-by-pixel flux fraction from TOI-5646 in S23. The red grid is the SPOC aperture. The pixel scale is 21 arcsec pixel$^{-1}$. The five sources that most contribute to the aperture flux are highlighted in different colours. The disk areas scale with the emitted fluxes. {\it Bottom panel}: flux contributions to the SPOC aperture from the target and most contaminant stars. This plot was created through \texttt{TESS-cont} (\url{https://github.com/castro-gzlz/TESS-cont}).} 
\label{fig:TESS-cont}
\end{figure}

\begin{figure}
\centering
\includegraphics[width=7.2cm]{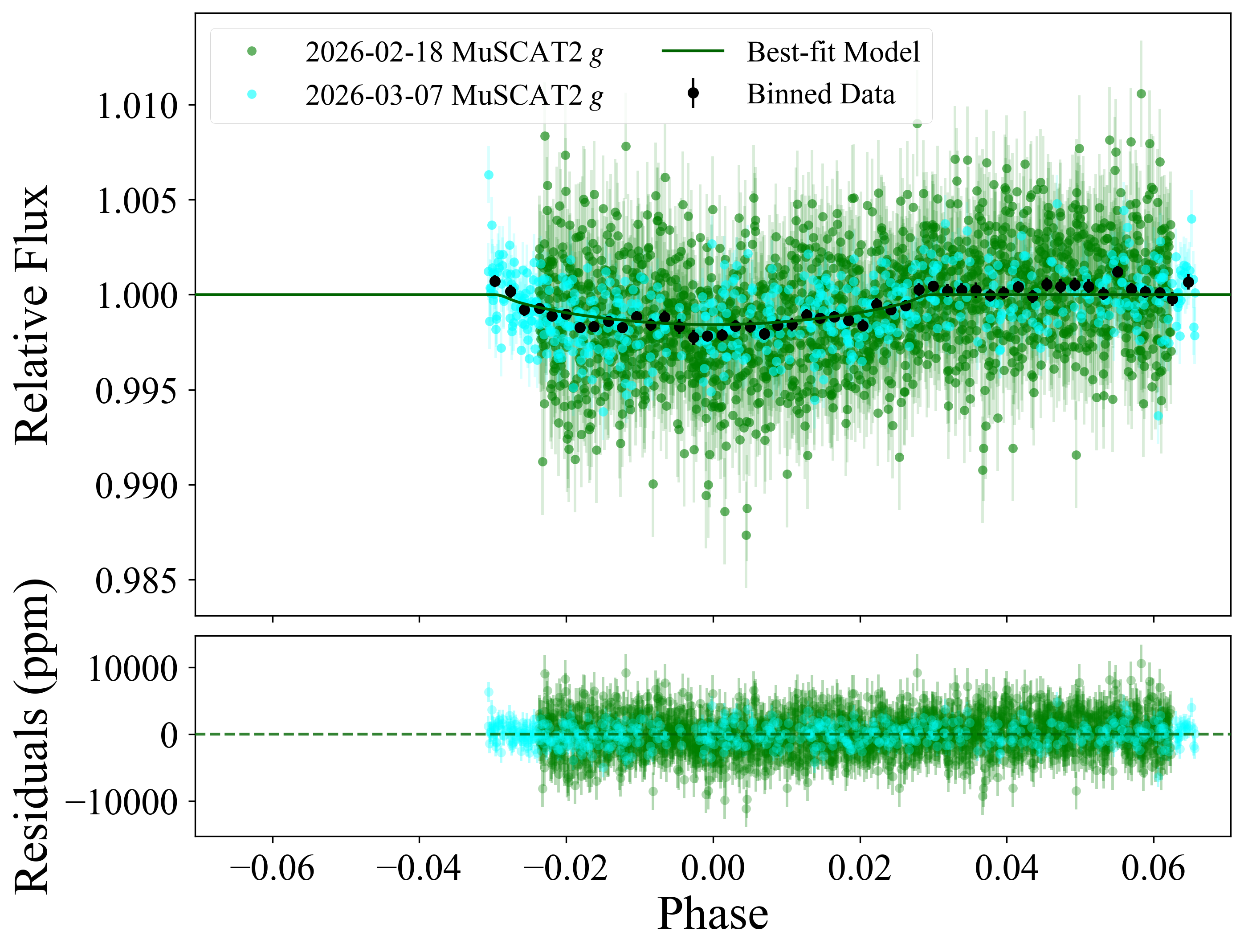}
\includegraphics[width=7.2cm]{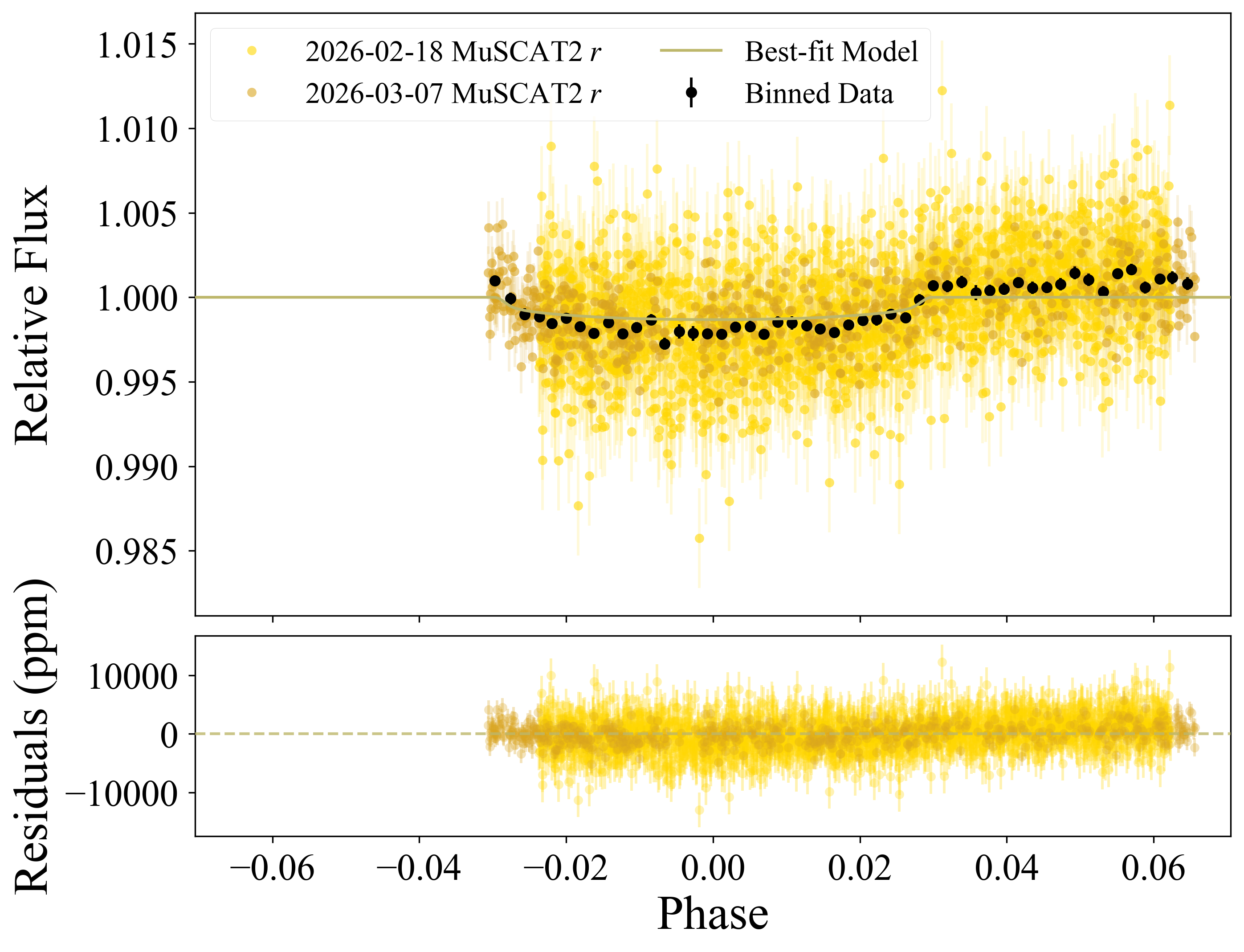}
\includegraphics[width=7.2cm]{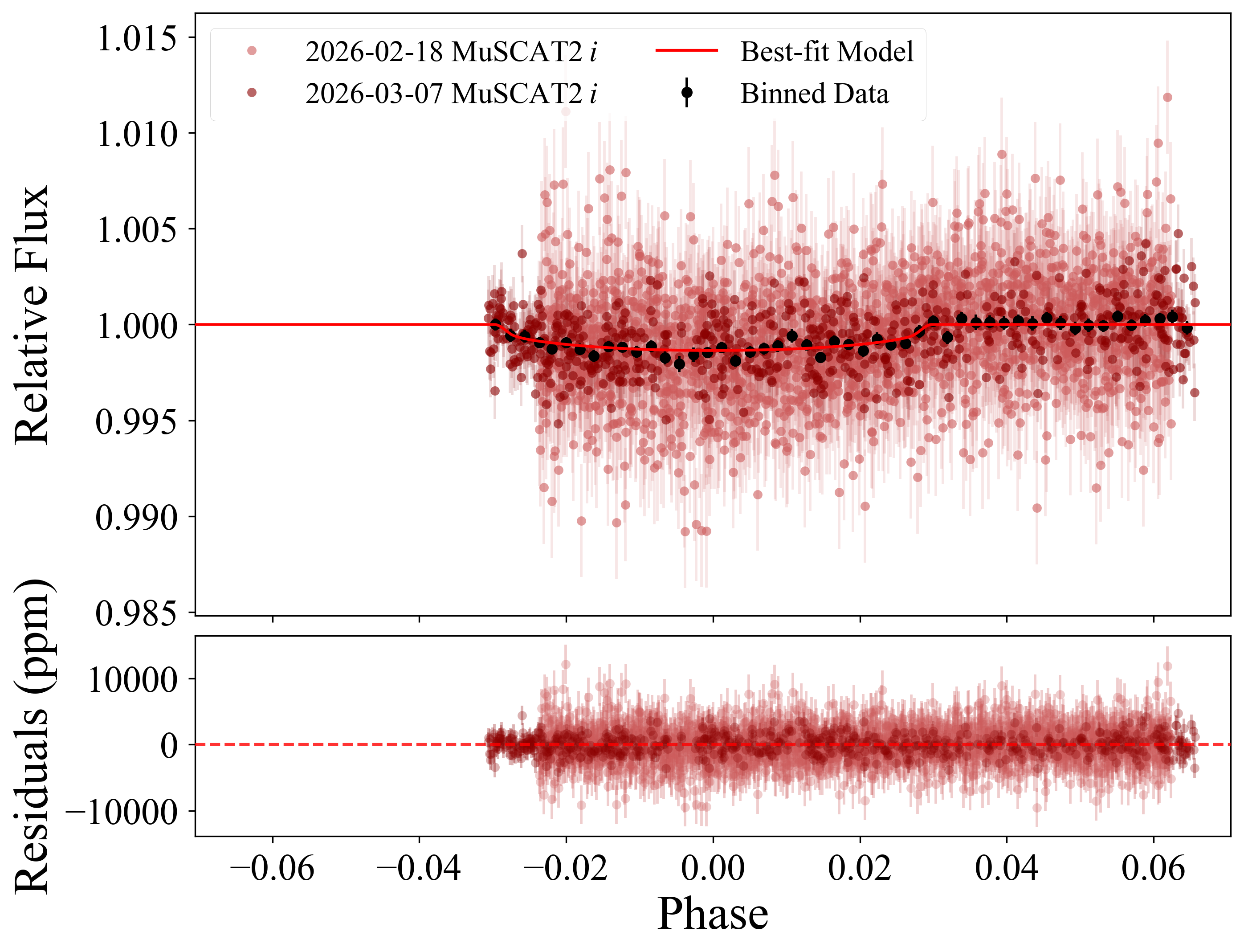}
\includegraphics[width=7.2cm]{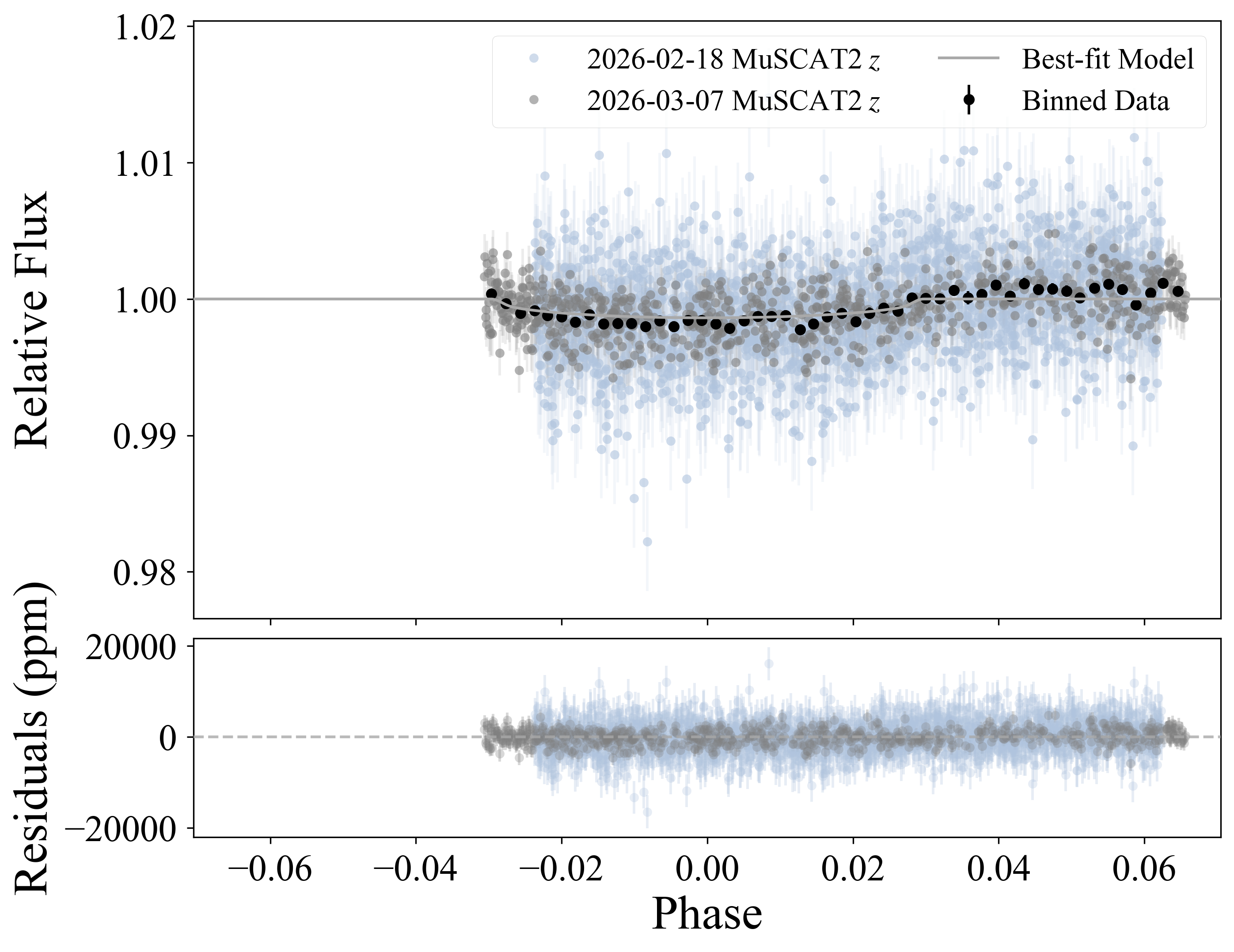}
\caption{Light curves of two transits of TOI-5646\,b observed with MuSCAT2. In each case, we show two panels ordered according to each of the four filters used; the top panel shows two phase-folded light curves observed with the same filter and the global fit (solid line), while the bottom panel shows residuals from the fits. The dark-filled circles show the light curves binned in phase.} 
\label{fig:muscat}
\end{figure}

\begin{figure}
\centering
\includegraphics[width=9.0cm]{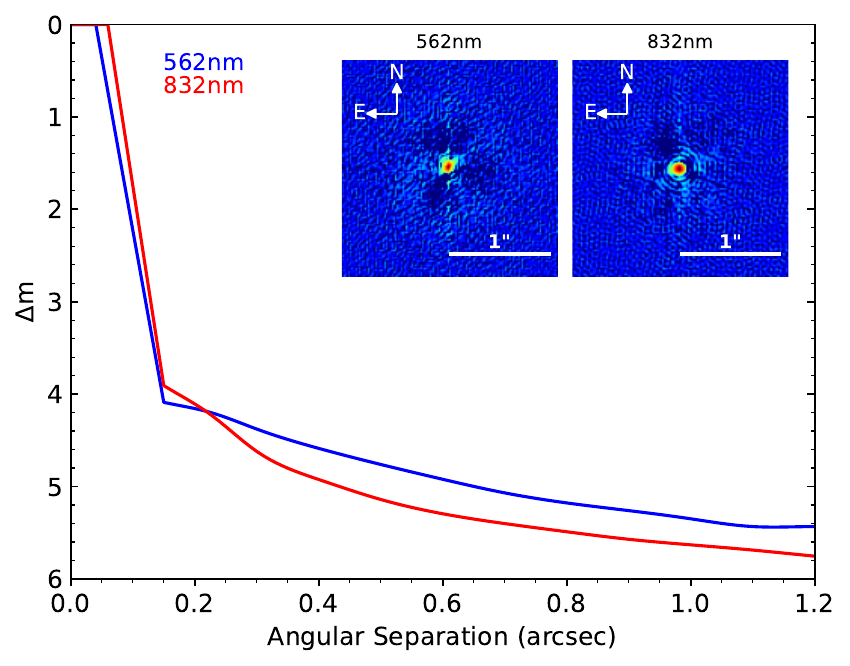}
\includegraphics[width=9.0cm]{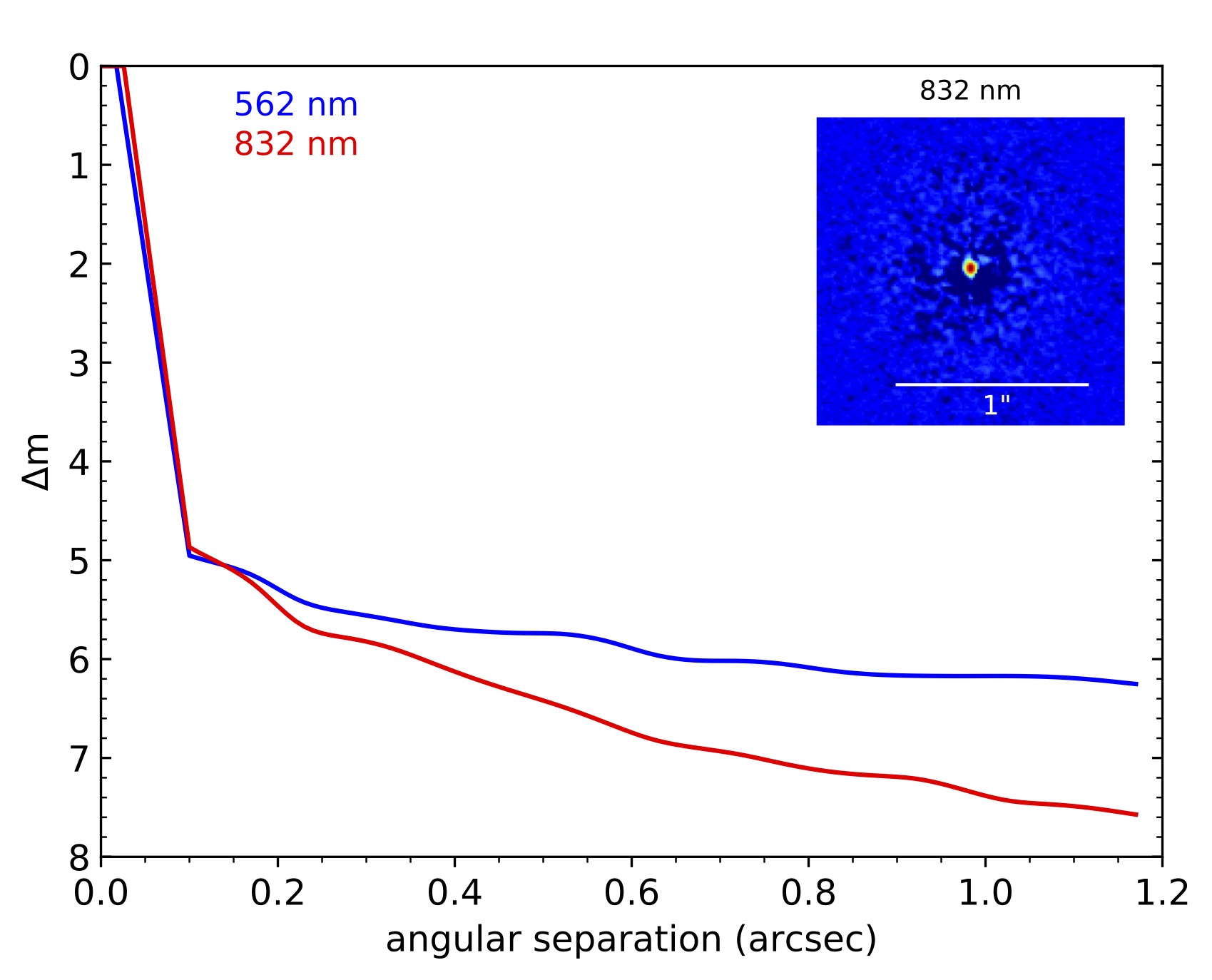}
\caption{High-resolution sensitivity curves. {\it Left-hand panel}: Contrast curves for TOI-5646 at 562~nm (blue) and 832~nm (red) obtained from NESSI speckle imaging. Reconstructed images are shown as insets to the plot with wavelengths labelled. {\it Right-hand panel}: 5\,$\sigma$-magnitude contrast curves in the two filters of ‘Alopeke as a function of the angular separation out to $1.2^{\prime \prime}$. The inset shows the reconstructed 832\,nm image of TOI-5646 with a $1^{\prime \prime}$ scale bar. TOI-5646 was found to have no close companions from the diffraction limit ($0.02^{\prime \prime}$) out to $1.2^{\prime \prime}$ to within the contrast levels achieved.} 
\label{fig:hri}
\end{figure}

\begin{figure}
\centering
\includegraphics[width=9.0cm]{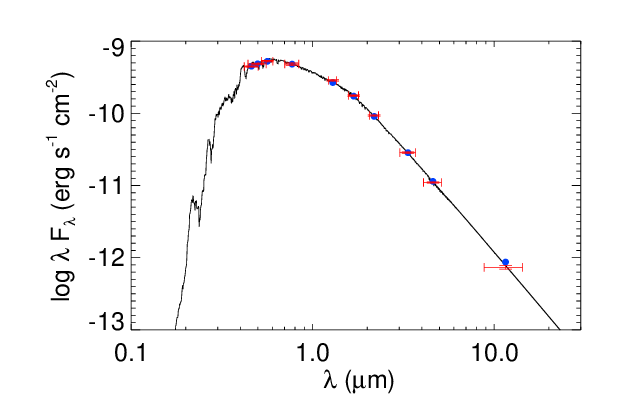}
\caption{Spectral energy distribution (SED) of the host star TOI-5646. The broad-band measurements from the APASS\,Johnson, APASS\,Sloan, 2MASS and WISE magnitudes are displayed in red, and the corresponding theoretical values with blue circles. The solid black line shows the non-averaged best-fit model.} 
\label{fig:SED}
\end{figure}

\begin{figure}
\centering
\includegraphics[width=9.0cm]{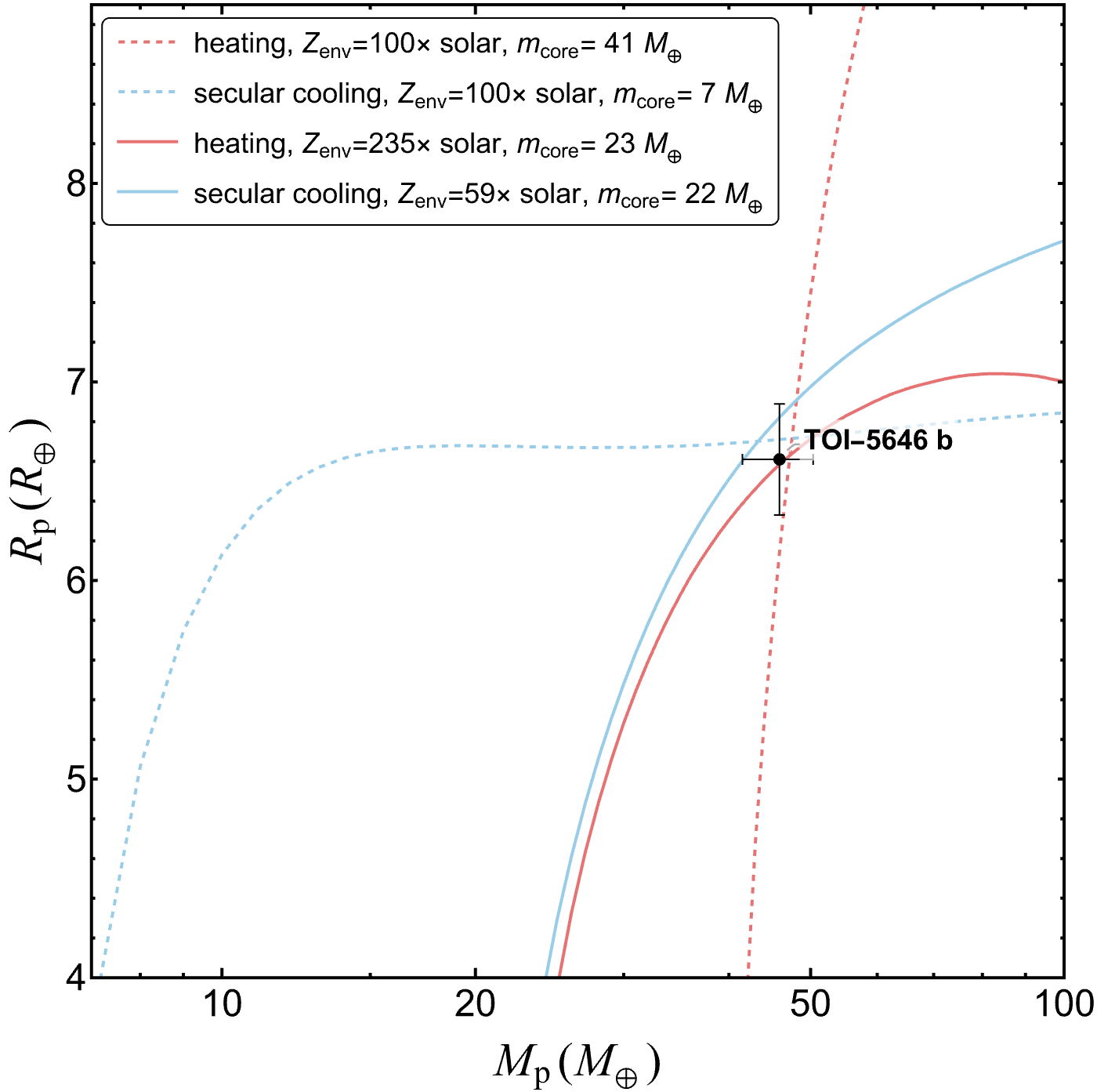}
\caption{Position of TOI-5646\,b in the radius-mass diagram. We display mass-radius curves for four possible interior structures of TOI-5646\,b (see text).} 
\label{fig:RpVsMp_models}
\end{figure}

\end{appendix}
\end{document}